# Unification of gravity and the harmonic oscillator on a quantum black hole horizon I: Nonperturbative particle scattering and Veneziano amplitudes

Marcia J. King

*228 Buckingham Avenue, Syracuse, NY 13210 and P. O. Box 1483, Borrego Springs, Ca 92004; e-mail: mking52701@aol.com*

(October 27, 2003)

A relativistic quantum model of particle scattering near the horizon of a microscopic black hole unifies gravity and the harmonic-oscillator force. The model is obtained by modifying a harmonic-oscillator nonstandard Lagrangian for a closed system of relativistic quarks to eliminate any cosmic background frame. Formulated in terms of the Planck units, the model has only one parameter, the cosmic number $N_2$. Other results include (1) quark confinement through cluster decomposition rather than a binding potential, (2) lepton substructure without violation of known experimental results, (3) a spontaneously broken symmetry between leptons and hadrons yielding leptons behaving as free particles and hadrons which scatter with Veneziano-type amplitudes, (5) realistic particle masses and sizes, and (6) a solution of a cosmic-number problem in cosmology.

## I. INTRODUCTION

Gerardus ‘t Hooft, emphasizing that near the horizon of a spherically symmetric static black hole the coordinates of the Schwarzschild solution can be transformed to Rindler coordinates in an almost flat space-time, has argued that it is reasonable to assume that the rules for quantum mechanics for elementary and composite particles should also hold there.[1-8] He further proposes that in such regions, string theory and gravity ought to be describable by a unified quantum theory. A major purpose of this article is to present a model which explicitly accomplishes these goals. The infinitesimal modification of a harmonic-oscillator nonstandard Lagrangian for a closed relativistic system of n quarks leads to the elimination of any cosmic background reference frame in flat space-time. Solutions for decoupled groups of quarks from the total system yield subsystems describing elementary particle scattering near the horizon of a microscopic black hole. In these solutions, gravity is unified with the harmonic-oscillator force. Quarks oscillate about space-time trajectories determined by gravity. Quantization is accomplished by canonical rules of quantum mechanics. Scattering amplitudes are the overlap between physical initial and final states, generating Veneziano-type scattering amplitudes in the case of hadron scattering. The S-matrix approach, i.e., perturbation theory, does not apply.

I have previously proposed similar models,[9,10,11,12] but the connection to gravity was not apparent to me at those times. The models of Refs. 11 and 12 differ from the present article in that, the particles become strings in a certain mathematical limit applied after matrix elements are calculated. In this limit, quarks oscillate with infinite frequency, representing closed strings, or tubes in space-time. It appears unnecessary to go to such a limit, however, and in the model of the present paper, the quarks oscillate with finite frequencies. The quarks, in today's cosmic time, have internal oscillations with amplitudes the size of a nucleon. Only in the earliest cosmic times are amplitudes on the order of a Planck length. All versions of the model are described in four-dimensional space-time, possible because the quarks are not *physically* extended objects.

There is a strong motivation for background elimination. The classical general theory of relativity is independent of a background reference frame, and has led to many researchers suggesting that this absence is necessary for extending a relativistic particle quantum theory to a cosmological quantum theory which includes gravity.[13] Among those emphasizing the necessity of finding away to do quantum physics without a background space-time geometry are J. B. Barbour and L. Smolin,[14] A. Ashtekar and J. Lewandowski,[15] and C. Rovelli.[16]

A second major purpose of this work is to propose that leptons as well as hadrons are composite particles. Substructures for both constitute a Higgs-type mechanism which yields mass. The (identical) leptons of the model, although composites, behave as if free particles, since they scatter in the forward or backward directions only. Hadron scattering, as mentioned above, is described by Veneziano-type amplitudes.

Another important objective of this article is to demonstrate that nonrelativistic quantum mechanics, under the proper circumstances, can be generalized to the relativistic regime in a very straightforward manner. The model of this article as well as the earlier versions has a particularly simple formalism based on a particle ontology and relativistic action at a distance, both of which have generally been thought to lead to unsurmountable difficulties (e.g., see Sec. XV). However, a combination of assumptions which separately fail yields a generalization of nonrelativistic quantum mechanics that is simple, realistic, self-consistent, and allows decoupling of groups of particles from the total system.

Relativistic quantum particle theories aside, there exists a cosmological motivation for adopting action at a distance. David Bohm's long exploration of nonlocality in quantum mechanics and his concept of implicate order have led to essays by many physicists contained in Ref. 17. The editors, B. J. Hiley and F. D. Peat, summarize Bohm's concept of implicate order, " . . . the relationship between two particles depends on something that goes beyond what can be described in terms of these two particles alone. In fact more generally, this relationship may depend on the quantum states of even larger systems, ultimately going on to the universe as a whole. Within this view separation becomes a contingent rather than a necessary feature of nature." As we shall see, the model stipulates the conditions for separation, i.e., cluster decoupling conditions, consequences of which are quark confinement, quantum gravity, and a Veneziano-type scattering amplitude for hadron scattering.

The approach here succeeds with the help of the simultaneous adoption of the following premises. The first premise, discussed in Sec. II, is parametric invariance of the action with a single (unmeasurable) evolution parameter separate from the space-time variables. In classical relativity, time is on an equal footing with space, but in nonrelativistic quantum mechanics, time t is treated as a c-number and separated from the dynamical variables. On the other hand, relativistic canonical methods of quantization are based on nonrelativistic quantum mechanics. This problem, known from the inception of quantum mechanics, is reviewed by K. Kuchar[18] concerning attempts to apply canonical methods, including Dirac's constraint method, in quantizing relativistic theories. Kuchar suggests putting the dynamical variables and time on an equal footing. Using a particle ontology, we show how a parametrically invariant action for a system of *interacting* quarks allows this without inconsistencies.

The second premise, discussed in Sec. III, is that natural boundary conditions[19,20] (n.b.c.) must be satisfied when the variational principle is applied to particle Lagrangians for which the canonical Hamiltonian vanishes. This natural assumption supplants any disregard of the boundary conditions[21] or other assumptions such as zero variation at the end points. Although the n.b.c. are Lorentz invariant, they are not form invariant under contact transformations, and are applied to "physical," not generalized, coordinates.

These premises allow the construction of a background-free relativistic quantum model for a universe of n quarks. Inputs are the Planck units $m_P$, $l_P$, $t_P$ and a cosmic number $N_2=(L_H/l_P)^{2/3}$ where $L_H$ Is the Hubble Length. The quarks are designated as"hadron" and "lepton" quarks for reasons that will become obvious. A brief outline of the article follows.

Spin and internal symmetries are at first omitted. A simple quark harmonic-oscillator model is constructed *with* a background frame. The Lagrangian is relativistic and nonstandard. Through the specification of initial conditions, it yields free composite particles (quark-antiquark pairs) obeying natural boundary conditions which include mass-shell relations and elimination of unwanted time oscillations. Solutions to the equations of motion correspond to either real or imaginary composite-particle mass, but the Dirac constraint $H \approx 0$ limits the allowable solutions to zero mass. This harmonic-oscillator model is next modified by adding to the coupling matrix a positive real infinitesimal parameter $\varepsilon$. This eliminates the background reference frame. It is demonstrated that for a set of decoupled quarks, this is equivalent to unifying the harmonic-oscillator force and gravity near the horizon of a black hole. Massless particles oscillate about trajectories determined by gravity in the flat Rindler space-time near the horizon of a spherically symmetric static black hole.

Since the foregoing describes massless composite particles, it does not yet represent a realistic model. The inclusion of two identical systems of quarks acts as a Higgs mechanism to obtain mass. These will be referred to as hadron and lepton quarks, respectively. Next, once again to eliminate the background frame, the coupling matrix is symmetrically modified by the addition of the parameter $\varepsilon$. This results in spontaneous symmetry breaking dictating different behaviors for the lepton and hadron quarks. Particle trajectories have solutions that are generalizations of those for the earlier massless particles, suggesting that they are related to trajectories in a Rindler space near horizons for other types of black holes.

Cluster decomposition is allowed under conditions that, for four-quark clusters, yield elastic composite-particle scattering and quark confinement. The cluster's space-time is not observable and acts like a black hole destroying and creating particles, as suggested by 't Hooft.[5] Hadron scattering is described by Veneziano-type amplitudes. Leptons scatter in the forward/backward directions only, which for these identical particles is tantamount to behaving like free particles. The non observable cluster space-times imply that the resulting composite particles have no memory of the interactions that produced them, making them suitable inputs for the perturbative model of Article II.

Squares of composite-particle masses and diameters are integral multiples of $N_2^{-1}\, m_P^2$ and $N_2\, l_P^2$, respectively, and thus change as the universe expands. It is argued that these simultaneous changes allow the electromagnetic coupling to remain constant.

Although $N_2$ increases as the observable universe expands, the cosmic-number relation[22] $N_1=N_2$ remains permanent. An observer detects no mass or size differences on his backward lightcone.

When spin and an SU(3) symmetry are added, hadrons split into three-quark baryons of half-integer spin, two-quark mesons of integer spin, and two-quark leptons with half-integer spin.

## II. SEPARATION OF THE EVOLUTION PARAMETER FROM KINEMATICS

The relativistic proper time for a single particle is defined by $d\tau$, where $d\tau^2 = g_{\mu\nu} dq^\mu dq^\nu / c^2$. The metric is $g_{00} = 1 = -g_{ii}$, $q^2 = g_{\mu\nu} q^\mu q^\nu$. The Poincaré and parametrically invariant action for a free particle is proportional to the proper time along the world line of the particle, or

$$I = -mc \int d\tau \left[\dot{q}^2(\tau)\right]^{1/2}, \tag{2.1}$$

where m has the dimension of mass.

The usual generalization of this action to a system of n free particles of equal mass is

$$I' = -mc \sum_{i=1}^{n} \int d\tau \left[\dot{q}_i^{\,2}(\tau)\right]^{1/2}. \tag{2.2}$$

Each term of the action remains proportional to a separate proper time along a particle's world line.

A variational method minimizing the integral of (2.2) leads to solutions

$$q_{i\mu} = a_{i\mu}\tau + b_{i\mu}, \quad i=1, 2, 3, \quad . . . , n. \tag{2.3}$$

Parametric invariance of the action yields n constraints on the conjugate momenta, known as the mass-shell conditions:

$$p_i^{\,2} = m^2 c^2. \tag{2.4}$$

The action (2.2) is not the only parametrically and Poincaré invariant action describing n particles with straight-line trajectories in space-time. Consider

$$I_n \equiv -mc \int ds \left\{ \sum_{i=1}^{n} \left[ \dot{q}_i^{\,2}(s) \right] \right\}^{1/2}. \tag{2.5}$$

The evolution parameter s is no longer associated with separate "proper times" since the action is not a sum of integrals. The single Dirac constraint establishes a relationship between the n particles:

$$\sum_{i=1}^{n} p_i^{\,2} = nm^2 c^2. \tag{2.6}$$

The action (2.5) sets our stage for the introduction of interactions. The simplest way to keep the action parametrically invariant while introducing the potential V is to write

$$I = -\int ds \left[ V(q_i) \sum_{i=1}^{n} \dot{q}_i^{\,2}(s) \right]^{1/2}. \tag{2.7}$$

Ref. 11 contains a discussion of the similarity of a harmonic-oscillator action to that for the free relativistic string,[23] namely,

$$I_s = -k \int d\tau \int_0^{\sigma_0} d\sigma \left[ -x'^2 \dot{x}^2 + (\dot{x} \cdot x')^2 \right]^{1/2}, \tag{2.8}$$

where k is a constant, $\dot{x}_\mu(\tau,\sigma) = dx_\mu / d\tau$, and $x'_\mu = dx_\mu / d\sigma$. In the present paper, we will strengthen the analogy by adding a second term in the square-root Lagrangian in (2.7).

## III. VARIATIONAL PRINCIPLE FOR A NONSTANDARD LAGRANGIAN AND NATURAL BOUNDARY CONDITIONS

Consider a system of n particles with coordinates $q_i(s)$ and velocities $\dot{q}_i(s) = dq_i(s)/ds$, with i = 1, 2, . . . , n. The variable s is the evolution parameter in units of time. Assume the Lagrangian is a known function of $q_i(s)$ and $\dot{q}_i(s)$ and is nonstandard. That is, the Lagrangian has the form

$$L\left(q_i(s),\dot{q}_i(s)\right) = \sum_{i=1}^{n} p_i(s)\,\dot{q}_i(s), \tag{3.1}$$

where $p_i \equiv \partial L/\partial \dot{q}_i$, and the canonical Hamiltonian vanishes.

C. Lanczos[20] notes that "Leibniz . . . advocated a quantity, the *vis viva* (living force) as the proper gauge for the dynamical action of a force. This *vis viva* of Leibniz coincides, apart from the unessential factor 2, with the quantity we call today 'kinetic energy.' " The Lagrangian above can be identified with this *vis viva*.

The action functional is defined as

$$I[C] = \int_{s_1}^{s_2} ds\, L\left(q_i(s),\dot{q}_i(s)\right) = \int_{s_1}^{s_2} ds \sum_{i=1}^{n} p_i(s)\,\dot{q}_i(s)\,. \tag{3.2}$$

Following Euler and Lagrange,[20,24] we assume a path exists in configuration space such that the action has a minimum value, and that this is the path chosen by *physical* (directly measurable) particles.

Therefore, we will consider a path $C'$ differing little from C and calculate the variation

$$\Delta I \equiv I[C'] - I[C]. \tag{3.3}$$

The $\Delta$ variation needs to be specified. In view of (3.1) and (3.2), we shall adopt the same variation as used for another action associated with the *standard* Lagrangian, namely,

$$A \equiv \int_{s_1}^{s_2} \sum_{i=1}^{n} p_i(s)\,\dot{q}_i(s) ds\,. \tag{3.4}$$

The variational principle applied to this action is sometimes called *the principle of Least Action*.[24] Displacements of $q_i(s)$ and $\dot{q}_i(s)$ are affected only by the "speeding up" or "slowing down" of each variable as a function of s, leaving the varied path consistent with the physical motion. The $\Delta$-process includes a variation of s at the end points, but the variations of the $q_i$'s remain zero. We consider the variations

$$q_i'(s) \equiv q_i(s) + \delta_i q_i(s)\,, \tag{3.5}$$

where

$$\delta_i q_i(s) \simeq \dot{q}_i(s)\,\delta_i s\,. \tag{3.6}$$

We have also

$$\dot{q}_i'(s) \equiv \dot{q}_i(s) + \frac{d}{ds}\delta_i\, q_i(s)\,. \tag{3.7}$$

Now compute the first-order change in the action functional I in going from C to $C'$:

$$\Delta I \equiv I[C'] - I[C] = \int_{s_1}^{s_2} L\left(q_i'(s),\dot{q}_i'(s)\right)ds - \int_{s_1}^{s_2} L\left(q_i(s),\dot{q}_i(s)\right)ds$$

$$= \int_{s_1}^{s_2} ds \sum_i \left[\frac{\partial L}{\partial q_i}\delta q_i(s) + \frac{\partial L}{\partial \dot{q}_i}\frac{d}{ds}\delta q_i(s)\right]$$

$$= \int_{s_1}^{s_2} ds \left[\sum_i \frac{\partial L}{\partial q_i} - \frac{d}{ds}\left(\frac{\partial L}{\partial \dot{q}_i}\right)\right]\dot{q}_i(s)\ \delta_i s + \sum_i \frac{\partial L}{\partial \dot{q}_i}\dot{q}_i(s)\ \delta_i s\ \Big|_{s_1}^{s_2}. \qquad (3.8)$$

Assume that to minimize the action integral, the total variation vanishes,[21] i.e.,

$$\Delta I = 0. \qquad (3.9)$$

The variations $\delta_i s$ are taken to be arbitrary and independent, and the end points $s_1$ and $s_2$ are arbitrary as well. For nontrivial solutions (where the $\dot{q}_i(s)$ are not identically zero), it follows that

$$\frac{\partial L}{\partial q_i} - \frac{d}{ds}\left(\frac{\partial L}{\partial \dot{q}_i}\right) = 0, \qquad (3.10)$$

and

$$p_i(s)\ \dot{q}_i(s)\big|_{s_1} = 0, \qquad p_i(s)\ \dot{q}_i(s)\ \big|_{s_2} = 0. \qquad (3.11)$$

The last conditions are the natural boundary conditions.[19,20] We denote them as the n.b.c.

The form of the Lagrange equations (3.10) is invariant under contact transformations, i.e., canonical transformations to a new set of coordinates $Q_i$ and $P_i$ such that

$$\sum_i P_i dQ_i = \sum_i p_i dq_i. \qquad (3.12)$$

However, the n.b.c. are not form invariant under such transformations. The variational principle selects out special coordinate systems. That is, it is based on a particle ontology, or, more specifically, the *vis viva* of Leibniz. The n.b.c. are invariant under the Lorentz transformations.

A problem of self-consistency arises if the n.b.c. are applied to the results of the last section. We resolve this by assuming that there are no free particles in nature. A potential must be introduced.

## IV. UNIFICATION OF THE HARMONIC-OSCILLATOR AND GRAVITY IN A SIMPLE QUANTUM QUARK MODEL

### A. Simple harmonic-oscillator model with a nonstandard Lagrangian

The harmonic-oscillator potential has appeared in various nonrelativistic quark models of hadrons, and was extended to include relativistic calculations and comparison to experiment in the well-known article by R. P. Feynman, M. Kislinger, and F. Ravndal.[25] Although problems such as time-oscillations prevented the quark model from being rigorously extended to the relativistic regime, this work has been an important motivation for the present model.

Consider a universe of 2N quarks with coordinates $Q_{IA\mu}(s)$ and $\dot{Q}_{IA\mu}(s) \equiv dQ_{IA\mu}(s)/ds$, where I= 1, 2, . . . , N, and A= 1, 2 (the choice of index notation differs from that in earlier articles[11,12] ). The parameter s has the dimension of time, but is not associated with any particle's proper time, as discussed in Sec. II. In Sec. II, we drew attention to the similarity between the harmonic-oscillator nonstandard Lagrangian and the relativistic string. Here we extend the similarity by adding a second term in the square-root Lagrangian, writing

$$L(s) = -\frac{c^3}{GN_2}\left[-\left(\mathbf{G}_2\mathbf{Q}(s)\right)^2\dot{\mathbf{Q}}^2(s) + \left(\dot{\mathbf{Q}}^T(s)\cdot\mathbf{G}_2\mathbf{Q}(s)\right)^2\right]^{1/2}. \tag{4.1}$$

Actually, the addition of the second term does not affect the results of the model in this article. However, we include it to leave the form of the Lagrangian model suitable for generalization and applicability under other circumstances. An example is considered in Article II.

The 2Nx2N coupling matrix $\mathbf{G}_2$ is dimensionless, and $N_2$ is the cosmic number defined by $N_2=(L_H/l_P)^{2/3}$ where $L_H$ is the Hubble length and $l_P$ is the Planck length. The Lagrangian describes action at a distance between the 2N quarks in the universe, while the observable universe has fewer quarks. The fixed parameters of the Lagrangian are the gravitational constant G, the speed of light c, and $N_2$. Dirac's constraint procedure[26] is used to define a Hamiltonian, and the Dirac gauge introduces the Planck constant h.

Note that the factor $(c^3/G)$ is equal to $m_P\,\omega_P$, where $m_P$ is the Planck mass, and $\omega_P$ is the inverse of the Planck time $t_P$. Recall that the Planck units for mass, length and time are

$$m_P = (\hbar c/G)^{1/2}, \quad l_P = (\hbar\, G/c^3)^{1/2}, \quad t_P = c^{-1}(\hbar\, G/c^3)^{1/2}. \tag{4.2}$$

The vector $\mathbf{Q}(s)$ has components $Q_{IA\mu}(s)$. The coupling matrix $\mathbf{G}_2$ is defined as

$$\mathbf{G}_2 \equiv \mathbf{g}\otimes\mathbf{N}, \tag{4.3}$$

with the $2\times2$ matrix $\mathbf{g}$ and the $N\times N$ matrix $\mathbf{N}$ given below:

$$\mathbf{g} \equiv \frac{1}{2}\begin{pmatrix} g_{11} & g_{12} \\ g_{12} & g_{11} \end{pmatrix}, \text{ and } \mathbf{N} \equiv \frac{1}{N}\begin{pmatrix} 1 & 1 & . & . & 1 \\ 1 & 1 & . & . & 1 \\ . & . & . & . & . \\ . & . & . & . & . \\ 1 & 1 & . & . & 1 \end{pmatrix} = \mathbf{N}^2. \tag{4.4}$$

We shall examine the case when $\mathbf{g}=\mathbf{g}^2$, which, besides the unit matrix, has the solutions

$$\mathbf{g}_2 = \frac{1}{2}\begin{pmatrix} 1 & -1 \\ -1 & 1 \end{pmatrix}, \text{and } \overline{\mathbf{g}}_2 = \frac{1}{2}\begin{pmatrix} 1 & 1 \\ 1 & 1 \end{pmatrix}. \tag{4.5}$$

Note that $\mathbf{g}_2 + \overline{\mathbf{g}}_2 = \mathbf{1}$, and $\mathbf{g}_2 \overline{\mathbf{g}_2} = 0$. Also, $\mathbf{G}_2 = \mathbf{G}_2{}^2$. We shall use $\mathbf{g}_2$.

For convenience, define

$$m(N_2) \equiv N_2{}^{-1/2} m_P, \quad \omega(N_2) \equiv N_2{}^{-1/2}\omega_P, \quad l(N_2) \equiv N_2{}^{1/2} l_P. \tag{4.6}$$

The nonstandard Lagrangian yields two primary constraints:

$$\Phi_1 \equiv \mathbf{P}^2 + m^2\omega^2\left(\mathbf{G}_2\mathbf{Q}\right)^2 \approx 0, \qquad \Phi_2 \equiv \mathbf{P}^T \cdot \mathbf{G}_2\mathbf{Q} \approx 0, \tag{4.7}$$

and a secondary constraint

$$\Phi_3 \equiv \mathbf{P}^T \cdot \mathbf{G}_2\mathbf{P} - m^2\omega^2\mathbf{Q}^T \cdot \mathbf{G}_2\mathbf{Q} \approx 0, \tag{4.8}$$

where $\mathbf{P}$ is the momentum conjugate to $\mathbf{Q}$.

Express the Dirac Hamiltonian as

$$H = \sum_{i=1}^{3} v_i \Phi_i \approx 0. \tag{4.9}$$

This constraint implies that the quantum state-vectors do not evolve in s.

Self consistency is maintained if we set $v_2 = v_3 = 0$. We choose a gauge factor

$$v_1 = \frac{1}{2m}, \tag{4.10}$$

where $m(N_2)$ introduces the Planck constant h. The harmonic-oscillator Hamiltonian is now

$$H = \frac{1}{2m}\left[\, \mathbf{P}^2 + m^2\omega^2\left(\mathbf{G}_2\mathbf{Q}\right)^2 \,\right]. \tag{4.11}$$

The matrix $\mathbf{g_2}$ can be diagonalized by

$$\delta_2 \equiv \frac{1}{\sqrt{2}}\begin{pmatrix} 1 & 1 \\ 1 & -1 \end{pmatrix} = \delta_2{}^{-1} \text{ and } \overline{\delta}_2 \equiv \frac{1}{\sqrt{2}}\begin{pmatrix} -1 & 1 \\ 1 & 1 \end{pmatrix} = \overline{\delta}_2{}^{-1}. \tag{4.12}$$

Choosing $\delta_2$, we define a transformation to a set of 2N coordinates:

$$Q_{IA} \equiv y_{IA} + N^{-1/2}\{\delta_2 \mathbf{W}\}_A, \qquad \text{with } \sum_{I=1}^{N} y_{IA} = 0. \tag{4.13}$$

The constraint on the $y_{IA}$ brings the number of independent coordinates back to 2N. In terms of the transformed coordinates, the conjugate momentum becomes

$$P_{IA} = p_{IA} + N^{-1/2}\{\delta_2\, \mathbf{p}\}_A , \tag{4.14}$$

where

$$p_{IA} \equiv m\, \dot{y}_{IA}, \quad \text{with} \quad \sum_{I=1}^{N} p_{IA} = 0; \qquad p_A \equiv m\dot{W}_A . \tag{4.15}$$

The center-of-mass vector is proportional to $W_1$ since

$$\frac{1}{N}\sum_{I=1}^{N}\sum_{A=1,2} Q_{IA} \equiv N^{-1/2}\sum_{A=1,2}\{\delta_2\, \mathbf{W}\}_A = \sqrt{2/N}\, W_1 . \tag{4.16}$$

The Hamiltonian can now be expressed as

$$H = \frac{1}{2m}\left[\sum_{I=1}^{N}\sum_{A=1,2} p_{IA}{}^2 + p_1{}^2 + \left(p_2{}^2 + m^2\omega^2 W_2{}^2\right)\right], \tag{4.17}$$

yielding equations of motion

$$\ddot{y}_{IA} = 0; \quad \ddot{W}_1 = 0; \quad \ddot{W}_2 = -\,\omega^2 W_2 . \tag{4.18}$$

Quantize by imposing the commutation relations $\left[Q_{IA\mu}, P_{IA\nu}\right] = -\,i\hbar\, g_{\mu\nu}$, or

$$\left[y_{IA\mu}, p_{IA\nu}\right] = -\,i\hbar\, g_{\mu\nu}, \quad I = 1,2,3,\ldots,N-1;$$

$$\left[W_{A\mu}, p_{A\nu}\right] = -\,i\hbar\, 2N g_{\mu\nu}, \quad A = 1,2 . \tag{4.19}$$

In the Heisenberg Picture, the solutions for $y_{IA}$ and $W_1$ are linear in s. Choose initial conditions for the solutions for $W_2$ and $p_2$ such that

$$W_2 = \sqrt{2N}\,(\hbar/2m\omega)^{1/2}\left[a_2{}^{\dagger}\exp(i\omega s) + a_2\exp(-\,i\omega s)\right],$$

$$p_2 = i\sqrt{2N}\,(\hbar m\omega/2)^{1/2}\left[a_2{}^{\dagger}\exp(i\omega s) - a_2\exp(-\,i\omega s)\right]. \tag{4.20}$$

The dimensionless harmonic-oscillator operators obey

$$\left[a_{2\mu}, a^{\dagger}{}_{2\nu}\right] = -\,g_{\mu\nu} . \tag{4.21}$$

For the nonstandard Lagrangian of this model, the Dirac constraint procedure implies that operators representing observables must commute with the constraint functions $\Phi_i$ and are therefore constant in s. This implies that the parameter s cannot be measured. The operators $p_{IA}$, $p_1$, and $n_2 \equiv -\,a_2{}^{\dagger}\cdot a_2$ represent simultaneously observable quantities, while $y_{IA}$, $W_1$, $W_2$, and $p_2$ are not observable. Note that this implies the quark coordinates $Q_{IA}$ and $P_{IA}$ are also not observable.

The Hamiltonian can now be expressed as

$$H = \frac{1}{2m}\left[\sum_{I=1}^{N}\sum_{A=1,2} p_{IA}{}^2 + p_1{}^2 - 4N\hbar m\omega \left(n_2 + 2\right)\right], \quad n_2 \equiv - a_2{}^{\dagger}\cdot a_2 . \tag{4.22}$$

Quark solutions are

$$Q_{IA} = y_{IA} + (2N)^{-1/2} W_1 \pm \frac{1}{\sqrt{2}} l \left[a_2{}^{\dagger}\exp(i\omega s) + a_2 \exp(-i\omega s)\right];$$

$$P_{IA} = p_{IA} + (2N)^{-1/2} p_1 \pm \frac{i}{\sqrt{2}} mc \left[a_2{}^{\dagger}\exp(i\omega s) - a_2 \exp(-i\omega s)\right]. \tag{4.23}$$

Clearly quarks can be paired into composite particles by boundary or initial conditions. For example, setting $y_{I1}=y_{I2}$ yields free composites described by

$$Q_I \equiv \tfrac{1}{2}\left(Q_{I1} + Q_{I2}\right) = y_{I1} + (2N)^{-1/2} W_1 ; \quad P_I \equiv \left(P_{I1} + P_{I2}\right) = 2p_{I1} + 2(2N)^{-1/2} p_1 . \tag{4.24}$$

The internal states of these composites are all described by the same function

$$q \equiv \left(Q_{I1} - Q_{I2}\right) = \pm 2l \left[a_2{}^{\dagger}\exp(i\omega s) + a_2 \exp(-i\omega s)\right]. \tag{4.25}$$

Application of the variational principle to the Lagrangian yields a set of natural boundary conditions (n.b.c.), which can be re-expressed in terms of the quark momenta as

$$P_{IA}{}^2(s) \sim 0 \quad \text{as } s \to \pm\infty . \tag{4.26}$$

This looks like a Klein-Gordon equation for a zero-mass quark. However, the n.b.c. yield

$$\left(p_{IA} + \frac{1}{\sqrt{2N}} p_1\right)^2 |\Psi\rangle = m^2c^2\left(n_2 + 2\right)|\Psi\rangle ,$$

$$a_2{}^2|\Psi\rangle = 0, \qquad \left[p_{IA} + \frac{1}{\sqrt{2N}} p_1\right]\cdot a_2 |\Psi\rangle = 0, \quad A = 1,2 . \tag{4.27}$$

Thus, a mass and a Klein-Gordon equation is associated with each composite particle.

Each solution yields a common mass for all two-quark composites, namely,

$$m_{n_2} = 2m\sqrt{n_2 + 2} , \tag{4.28}$$

and the common diameter

$$\left|q_{n_2}\right| = 2l\sqrt{n_2 + 2} . \tag{4.29}$$

The n.b.c. imply that for a positive mass-squared composite particle in its rest frame

$$a_{20}|\Psi\rangle = a_{20}\left|p_{IA}, p_1, n_2\right\rangle = 0 . \tag{4.30}$$

In other words, there are no time oscillations in the rest frame of positive mass-squared particles.

Rephrasing the constraints $\Phi_i$, I=1, 2, 3, in terms of quantized operators, we find that the n.b.c.

imply $\Phi_2$ and $\Phi_3$ vanish (the converse is not true). The first constraint yields the Dirac Hamiltonian constraint $H \approx 0$, which, with the mass-shell constraints, implies

$$\left(n_2 + 2\right) = 0. \tag{4.31}$$

We show in Sec. V how this zero-mass problem can be resolved by a Higgs-type mechanism through the introduction of imaginary masses.

If we substitute the matrix $\overline{\mathbf{g}}_2$ for $\mathbf{g}_2$, and/or $\overline{\delta}_2$ for $\delta_2$, we get equivalent solutions.

## B. Unification of gravity and the harmonic-oscillator force near a black-hole horizon

To eliminate the background reference frame for this simple model, we replace the coupling matrix $\mathbf{G_2}$ by a new coupling matrix

$$\mathbf{H_2} \equiv \left(\mathbf{g_2} - i\varepsilon\,\overline{\mathbf{g}}_\mathbf{2}\right) \otimes \mathbf{N} + i\varepsilon \mathbf{1} \otimes \mathbf{1}, \tag{4.32}$$

where $\varepsilon$ is a positive infinitesimal real parameter. The Dirac Hamiltonian now becomes

$$H = \frac{1}{2m}\left[\mathbf{P}^2 + m^2\omega^2(\mathbf{H_2 Q})^2\right]. \tag{4.33}$$

Transformations (4.13) and (4.14) yield

$$H = \frac{1}{2m}\left[\sum_{!,A}\left(p_{IA}{}^2 - (m\omega\varepsilon)^2 y_{IA}{}^2\right) + p_1{}^2\left(p_2{}^2 + (1 + i\varepsilon)m^2\omega^2 W_2\right)^2\right]. \tag{4.34}$$

$W_1$ is the only variable which is linear in s. Express the remaining solutions as

$$y_{IA} = (\varepsilon m\omega)^{-1}\left[a_{IA}\exp(\varepsilon\omega s) + b_{IA}\exp(-\varepsilon\omega s)\right],$$

$$p_{IA} = \left[a_{IA}\exp\left(\varepsilon_N\omega s\right) - b_{IA}\exp\left(-\varepsilon_N\omega s\right)\right];;$$

$$W_2 = \left(\frac{N\hbar}{m\omega}\right)^{1/2}\left[a_2{}^\dagger\exp(i(1+i\varepsilon)\omega s) + a_2\exp(-i(1+i\varepsilon)\omega s)\right],$$

$$p_A = i\,(N\hbar m\omega)^{1/2}(1 + i\varepsilon)\left[a_2{}^\dagger\exp(i(1+i\varepsilon)\omega s) - a_2\exp(-i(1+i\varepsilon)\omega s)\right]. \tag{4.35}$$

Imposing the quantum conditions $\left[Q_{IA\mu}, P_{IA\nu}\right] = -i\hbar\, g_{\mu\nu}$ yields

$$\left[y_{IA\mu}, p_{IA\nu}\right] = -i\hbar\, g_{\mu\nu},\ I = 1,2,3,\ldots,N-1; \qquad \left[W_{A\mu}, p_{A\nu}\right] = -i\hbar\, 2Ng_{\mu\nu}, \tag{4.36}$$

and in turn, to first order,

$$\left[a_{IA\mu}, b_{IA\nu}\right] = 0; \quad \left[a_{2\mu}, a_{2\nu}{}^\dagger\right] = -g_{\mu\nu}. \tag{4.37}$$

The variables $W_1$ and $p_1$ are proportional to the total center-of-mass and the total momentum, respectively. From the above solutions, we see that composite particles can be formed only in the asymptotic regions of s where $W_1$ and $p_1$ play no role. However, the constant $p_1$ appears in the constraint $H \approx 0$. Assuming the total momentum of the system is zero or finite and N large, we shall drop the term $p_1^{\,2}$ in the Hamiltonian. We have, then, to first order,

$$H = \frac{-2}{m}\left[\sum_{I=1}^{N}\sum_{A=1}^{4} a_{IA}\cdot b_{IA} + Nm^2c^2\left(n_2 + 2\right)\right]. \tag{4.38}$$

The solutions for $W_2$ and $p_2$ are complex numbers. If we define the position and momentum variables as $(W_2+W_2^{\dagger})$ and $(p_2+p_2^{\dagger})$, then as $s \to \pm\infty$,

$$\left(W_2 + W_2^{\dagger}\right) \sim \left(\frac{N\hbar}{m\omega}\right)^{1/2} \exp|\varepsilon\omega s|\left[a_2^{\dagger}\exp(i\omega s) + a_2\exp(-i\omega s)\right], \tag{4.39}$$

$$\left(p_2 + p_2^{\dagger}\right) \cong i\,(N\hbar m\omega)^{1/2}\exp|\varepsilon\omega s|\left[a_2^{\dagger}\exp(i\omega s) - a_2\exp(-i\omega s)\right]. \tag{4.40}$$

Since $W_2 + W_2^{\dagger} \simeq W_2$ and $p_2 + p_2^{\dagger} \simeq p_2$ as $s \to \pm\infty$ and $\varepsilon \to 0$, we shall retain the notation $W_2$ and $p_2$ with the understanding that we mean $W_2+W_2^{\dagger}$ and $p_2+p_2^{\dagger}$.

To first order, commuting operators are $a_{IA}$, $b_{IA}$, and the number operator $n_2 \equiv -a_2^{\dagger}a_2$. Write the state vectors as

$$|\Psi\rangle \equiv \left|a_{IA}, b_{IA}, n_2\right\rangle. \tag{4.41}$$

The natural boundary conditions are $P_{IA}^{\,2}(s) \sim 0$ as $s \to \pm\infty$, and yield

$$a_{IA}^{\,2}|\Psi\rangle = b_{IA}^{\,2}|\Psi\rangle = m^2c^2\left(n_2 + 2\right)|\Psi\rangle, \tag{4.42}$$

$$a_2^{\,2}|\Psi\rangle = 0, \qquad a_{IA}\cdot a_2|\Psi\rangle = 0, \qquad b_{IA}\cdot a_2|\Psi\rangle = 0. \tag{4.43}$$

The n.b.c. lead to the satisfaction of the Dirac constraints other than $H \approx 0$.

The Lagrangian allows the straightforward consideration of cluster decomposition, which is necessary to describe the kinematically uncorrelated results of distant experiments. Note that the quark coordinates are independent of N. Consider a cluster of four quarks corresponding to I =1, 2. Kinematic decoupling is created by setting

$$\sum_{I=3}^{N} a_{IA} = \sum_{I=3}^{N} b_{IA} = 0, \qquad \sum_{I=3}^{N}\sum_{A=1,2} a_{IA}\cdot b_{IA} = 0. \tag{4.44}$$

It follows that

$$\sum_{I=1}^{2} a_{IA} = \sum_{I=1}^{2} b_{IA} = 0, \tag{4.45}$$

and

$$\sum_{I=1}^{2} \sum_{A=1,2} a_{IA} \cdot b_{IA} = 0. \tag{4.46}$$

It is important to note that if N=2 described the total system, the last condition would not follow. We shall consider initial states (observer time $t = -\infty$) that satisfy (1) cluster decomposition conditions, (2) the n.b.c. equations, (3) the constraint $H \approx 0$.

For the decoupled subgroup of four quarks, the quark position vectors are

$$Q_{IA} = \frac{1}{\varepsilon m \omega}\left[a_{IA} \exp(\varepsilon\omega s) + b_{IA} \exp(-\varepsilon\omega s)\right] + \frac{1}{\sqrt{2N}} W_1 + (-1)^{A-3} \sqrt{2}\, l \cosh(\varepsilon\omega s)\left[a_2^{\dagger} \exp(i\omega s) + a_2 \exp(-i\omega s)\right], \tag{4.47}$$

where $I = 1,2;\ A = 1,2$. From cluster decomposition, it follows that,

$$a_{11} = -a_{21},\ a_{12} = -a_{22},\ b_{11} = -b_{21},\ b_{12} = -b_{22},\ a_{11} \cdot b_{11} = -a_{12} \cdot b_{12}. \tag{4.48}$$

Consider the case

$$a_{110} < 0,\ b_{110} < 0,\ a_{220} < 0,\ b_{120} < 0. \tag{4.49}$$

The first two relations imply that the quark $Q_{11}$ turns around in time, so that at $t = -\infty$, the same quark appears in two different locations in space. The other quarks at $t = -\infty$ are $Q_{22}$ and $Q_{12}$.

Consider now what quarks exist at $t = +\infty$. It follows from the above that

$$a_{120} > 0,\ b_{220} > 0,\ a_{210} > 0,\ b_{210} > 0. \tag{4.50}$$

Thus, the quarks $Q_{22}$, $Q_{12}$, and $Q_{21}$ exist in the final state $t = +\infty$, with $Q_{21}$ at two spatial locations.

To describe a composite particle in the initial state, apply quark confinement , say $a_{11} = a_{22}$, as an initial condition at $t = -\infty$. We have then

$$a_{11} \cdot (b_{11} - b_{12}) = 0. \tag{4.51}$$

We shall assume that the choices of arbitrary constants at $s = -\infty$ are independent of the choices at $s = +\infty$. The above equation then implies $b_{11}=b_{12}$, or confinement for the other initial-state quarks. For the final state at $t = +\infty$, we find

$$a_{21} = a_{12} \text{ with } a_{210} = a_{120} > 0, \text{ and } b_{22} = b_{21} \text{ with } b_{220} = b_{210} > 0. \tag{4.52}$$

In other words, an initial state composite and the decoupling conditions imply that the other two quarks in the initial state and the four quarks in the final state are confined pairwise into composites as well.

Consider now the composites formed in the asymptotic regions $s \to \pm\infty$. With the aid of the quark solutions, we obtain

$$Q_1^+ \equiv \tfrac{1}{2}(Q_{11}+Q_{22})\Big|_{s\to+\infty} \sim \left(1/\varepsilon m\omega\right)\, a_{11}\exp(\varepsilon\omega s),$$

$$Q_2^+ \equiv \tfrac{1}{2}(Q_{21}+Q_{12})\Big|_{s\to+\infty} \sim -\left(1/\varepsilon m\omega\right)\, a_{11}\exp(\varepsilon\omega s),$$

$$Q_1^- \equiv \tfrac{1}{2}(Q_{11}+Q_{12})\Big|_{s\to-\infty} \sim \left(1/\varepsilon m\omega\right)\, b_{11}\exp(-\varepsilon\omega s),$$

$$Q_2^- \equiv \tfrac{1}{2}(Q_{21}+Q_{22})\Big|_{s\to-\infty} \sim -\left(1/\varepsilon m\omega\right)\, b_{11}\exp(-\varepsilon\omega s). \qquad (4.53)$$

The composites have oscillatory internal states described by

$$q(s) \equiv \sqrt{2}\, l \exp\left(|\varepsilon\omega s|\right)\left[a_2^{\dagger}\exp\left(i\omega s\right) + a_2\exp\left(-i\omega s\right)\right]. \qquad (4.54)$$

The composite momenta are

$$P_1^+ \equiv \left(P_{11}+P_{22}\right)\Big|_{s\to+\infty} \sim 2a_{11}\exp(\varepsilon\omega s),$$

$$P_2^+ \equiv \left(P_{21}+P_{12}\right)\Big|_{s\to+\infty} \sim -2\, a_{11}\exp(\varepsilon\omega s),$$

$$P_1^- \equiv \left(P_{11}+P_{12}\right)\Big|_{s\to-\infty} \sim -2\, b_{11}\exp(-\varepsilon\omega s),$$

$$P_2^- \equiv \left(P_{21}+P_{22}\right)\Big|_{s\to-\infty} \sim 2\, b_{11}\exp(-\varepsilon\omega s). \qquad (4.55)$$

Four-momentum and angular momentum are conserved. The n.b.c. imply

$$m_{n_2}{}^2 = 4m^2(n_2+2), \qquad (4.56)$$

but the constraint $H \approx 0$ implies that the composite masses must be zero.

All relevant quantities in the asymptotic regions $s \to \pm\infty$ are scaled by the common factor $\exp|\varepsilon\omega s|$. The smallest measuring units available are $l_\mu\left(N_2\right)\exp|\varepsilon\omega s|$, so that in these basic units, the coordinates and momenta appear constant. We can remove the common factor. Alternatively, if we let $\varepsilon \to 0$ after letting $s \to \pm\infty$, the momenta and position vectors are constant, but the position vectors are singular. We conclude that the cluster yields free composite particles with observable momenta, but its space-time is not defined as $\varepsilon \to 0$.

The following picture of scattering emerges. For $s \to \pm\infty$, zero-mass composites are formed in the asymptotic regions of the cluster's space-time. We take the initial (final) state of the particles in the observer's frame as the initial (final) state in the cluster space-time. Scattering occurs by quark exchange in the forward or backward directions only.

Return now to the general position vectors for the quarks in the cluster, which can be expressed as

$$Q_{IA} = q_{IA} + (-1)^{A-3}\sqrt{2}\, l \cosh(\varepsilon\omega s)\left[a_2^{\dagger}\exp(i\omega s) + a_2\exp(-i\omega s)\right], I = 1,2;\ A = 1,2. \quad (4.57)$$

The trajectories $q_{IA}$ can be taken to be in the z-t plane in a frame such that

$$q_{22,3} = \frac{1}{g}\cosh(gs) = -q_{12,3}, \quad (4.58)$$

$$q_{22,0} = -\frac{1}{g}\sinh(gs) = -q_{12,0}. \quad (4.59)$$

These are equations describing the trajectory of a particle of real mass under constant acceleration in Minkowski space.[27] The particle experiences a constant gravitational field. The position of the particle at the "time" s is obtained from the position at s=0 by a Lorentz transformation.

The remaining two trajectories, which turn around in time in the observer's frame, are

$$q_{11,3} = \frac{1}{g}\sinh(gs) = -q_{21,3}, \quad (4.60)$$

$$q_{11,0} = -\frac{1}{g}\cosh(gs) = -q_{21,0}, \quad (4.61)$$

which correspond to the Lorentz boost of an imaginary mass particle.

A real mass particle has velocity $0 \le v < c$, whereas an imaginary mass particle (tachyon) has velocity $c < v \le \infty$. For our cluster of four quarks, two quarks start with velocity c and are slowed down to zero velocity, while the other two start with velocity c and reach infinite velocity which allows them to turn around in the observer's time.

To relate the solutions to gravity near the horizon of a black hole, we briefly review Rindler space-time near the horizon of a static spherically symmetric black hole.[1,5] The Schwarzschild metric is

$$ds^2 = -(1 - \frac{2M}{r})dt^2 + (1 - \frac{2M}{r})^{-1}dr^2 + r^2 d\Omega^2, \quad (4.62)$$

with

$$d\Omega^2 = d\theta^2 + \sin^2\theta\, d\phi^2, \quad (4.63)$$

and M is a constant of integration. A transformation is made to the Kruskal coordinates $(x, y, \theta, \phi)$ defined through the relations

$$\left(\frac{r}{2M} - 1\right)e^{r/2M} \equiv xy, \qquad e^{t/2M} \equiv \frac{x}{y}. \quad (4.64)$$

This yields

$$ds^2 = \frac{32}{r} M^3 e^{-r/2M} dx\, dy + r^2 d\Omega. \tag{4.65}$$

Near the horizon, or $r \approx 2M$, the metric becomes

$$ds^2 \approx \frac{16M^2}{e} dx\, dy + 4M^2 d\Omega^2. \tag{4.66}$$

The Rindler coordinates X, Y, Z, T are defined by

$$Z + T \equiv \frac{4M}{\sqrt{e}} x, \quad Z - T \equiv \frac{4M}{\sqrt{e}} y, \quad X \equiv 2M(\theta - \frac{\pi}{2}), \quad Y \equiv 2M\phi. \tag{4.67}$$

Close to the origin in these coordinates, space-time is approximately flat, or

$$ds^2 \approx -dT^2 + dZ^2 + dX^2 + dY^2. \tag{4.68}$$

Consider the transformation

$$Z \equiv \rho \cosh\tau, \quad T \equiv \rho \sinh\tau. \tag{4.69}$$

By writing

$$t = 4M\,\tau, \quad 8M(r - 2M) = \rho^2, \tag{4.70}$$

we regain the Schwarzschild Metric near the horizon of the black hole. The equation (4.69) represents a Lorentz transformation. A spherically symmetric observer in the Rindler coordinates has $\rho$ constant, or feels a gravitational field which is constant in time. Apparently the approximately flat Rindler space-time is closely connected to the introduction of the parameter $\varepsilon$. We leave this for future analysis.

## V. A SUPERSYMMETRY: LEPTON AND HADRON QUARKS

The model of the last section is modified by including a second identical system of quarks. The total system of 4N quarks is described by coordinates $Q_{IA\mu}$ and $\dot{Q}_{IA\mu}(s)$, where I=1, 2, . . . , N, and A=1, 2, 3, 4. The action-at-a-distance Lagrangian is

$$L(s) = -m\omega\left[-(\mathbf{GQ}(s))^2\, \dot{\mathbf{Q}}^2(s) + \left(\dot{\mathbf{Q}}^T(s)\cdot \mathbf{GQ}(s)\right)^2\right]^{1/2}, \tag{5.1}$$

where m and $\omega$ are defined in the last section. The $4N \times 4N$ dimensionless coupling matrix $\mathbf{G}$ is taken to be

$$\mathbf{G} \equiv \mathbf{g}_4 \otimes \mathbf{N}, \tag{5.2}$$

with

$$\mathbf{g}_4 \equiv \begin{pmatrix} \mathbf{g}_2 & 0 \\ 0 & \mathbf{g}_2 \end{pmatrix} = \mathbf{g}_4{}^2 . \tag{5.3}$$

Denote the A=1 and A=2 quarks as hadron quarks, and the A=3 and A=4 quarks as lepton quarks.

The nonstandard Lagrangian yields Dirac constraints analogous to those of the last section. We choose the Dirac Hamiltonian

$$H = \frac{1}{2m} [\, \mathbf{P}^2 + m^2 \omega^2 (\mathbf{GQ})^2 \,] . \tag{5.4}$$

The matrix $\mathbf{g}$ can be diagonalized by either of the following matrices

$$\delta \equiv \frac{1}{\sqrt{2}} \begin{pmatrix} \delta_2 & 0 \\ 0 & \delta_2 \end{pmatrix} = \delta^{-1}, \qquad \rho \equiv \frac{1}{\sqrt{2}} \begin{pmatrix} -\delta_2 & \delta_2 \\ \delta_2 & \delta_2 \end{pmatrix} = \rho^{-1} . \tag{5.5}$$

The matrix $\delta$ maintains the symmetry between the lepton and hadron quarks while $\rho$ does not. We will make use of this important fact later.

Define a transformation to a new set of 4N coordinates:

$$Q_{IA} \equiv y_{IA} + N^{-1/2}\{\delta\, \mathbf{W}\}_A , \qquad \text{with } \sum_{I=1}^{N} y_{IA} = 0. \tag{5.6}$$

In terms of the transformed coordinates, the conjugate momentum becomes

$$P_{IA} = p_{IA} + N^{-1/2}\{\delta\, \mathbf{p}\}_A , \tag{5.7}$$

where

$$p_{IA} \equiv m\, \dot{y}_{IA}, \quad \text{with } \sum_{I=1}^{N} p_{IA} = 0; \qquad p_A \equiv m\dot{W}_A . \tag{5.8}$$

The Hamiltonian becomes

$$H = \frac{1}{2m} \left[ \sum_{I=1}^{N} \sum_{A=1}^{4} p_{IA}{}^2 + \sum_{A=1,3} p_A{}^2 + \sum_{A=2,4} \left( p_A{}^2 + m^2\omega^2 W_A{}^2 \right) \right] . \tag{5.9}$$

Quantize by putting

$$\left[ y_{IA\mu}, p_{IA\nu} \right] = -i\hbar\; g_{\mu\nu}, \quad I = 1,2,3,\ldots,N-1;$$

$$\left[ W_{A\mu}, p_{A\nu} \right] = -i\hbar\; 2N g_{\mu\nu}, \quad A = 1,2,3,4. \tag{5.10}$$

The solutions for $y_{IA}$ and $W_1$ and $W_3$ are linear in s. Express the remaining solutions as;

$$W_A = \sqrt{2N}\; (\hbar/2m\omega)^{1/2} \left[ a_A{}^\dagger \exp(i\omega s) + a_A \exp(-i\omega s) \right],$$

$$p_A = i\sqrt{2N}\; (\hbar m\omega/2)^{1/2} \left[ a_A{}^\dagger \exp(i\omega s) - a_A \exp(-i\omega s) \right], \quad A = 2,4. \tag{5.11}$$

Note that the center of mass vector for the system is now proportional to $W_1 + W_3$. Had we instead defined the transformation from $Q_{IA}$ to $y_{IA}$ and $W_A$ by replacing $\delta$ by $\rho$, the center of mass vector would be proportional to $W_3$. We will capitalize on this lepton-hadron asymmetry later.

The commutation relation for $W_A$ and $p_A$ implies

$$\left[a_{A\mu}, a^{\dagger}_{A\nu}\right] = -g_{\mu\nu}, \quad A = 2, 4. \tag{5.12}$$

The Hamiltonian can be expressed as

$$H = \frac{1}{2m}\left[\sum_{I=1}^{N}\sum_{A=1}^{4} p_{IA}^{\ 2} + \sum_{A=1,3} p_A^{\ 2} - 4N\hbar m\omega \sum_{A=2,4}\left(n_A + 2\right)\right], \tag{5.13}$$

with $n_A \equiv -a_A^{\ \dagger}\cdot a_A$. Pairs of quarks in either system become composite particles by initial conditions as discussed in Sec. IV. The n.b.c. imply composite mass-shell relations of the form

$$P^2 = 4m^2c^2\left(n_A + 2\right), \tag{5.14}$$

where P is the momentum of the composite.

The constraint $H \approx 0$ and the mass-shell constraints together yield

$$\left(n_2 + 2\right) = -\left(n_4 + 2\right). \tag{5.15}$$

In other words, the inclusion of lepton quarks corresponding to imaginary mass leptons allows the hadron composites to have real mass, and vice versa. The n.b.c. imply the satisfaction of the remaining constraints (the converse is not true).

With the diagonalizing matrix $\delta$, the symmetry between the A=1, 2 and the A=3, 4 coordinates is apparent. Let us look at the solutions for the alternative diagonalizing matrix $\rho$. Replace the system of 4N coordinates $Q_{IA}$ by generalized coordinates $x_{IA}$, writing

$$L(s) = -m\omega\left[-\left(\mathbf{G}\,\mathbf{x}(s)\right)^2\dot{\mathbf{x}}^2(s) + \left(\dot{\mathbf{x}}^T(s)\cdot\mathbf{G}\mathbf{x}(s)\right)^2\right]^{1/2}. \tag{5.16}$$

Define the transformation

$$x_{IA} \equiv y_{IA} + N^{-1/2}\{\rho\,\mathbf{W}\}_A, \qquad \text{with } \sum_{I=1}^{N} y_{IA} = 0. \tag{5.17}$$

The solutions for $y_{IA}$ and $W_A$ are essentially the ones of the last section. However, the center-of-mass vector is proportional to $W_3$ instead of $W_1+W_3$.

We now assume that the *physical* quark coordinates $Q_{IA}$, i.e., those to which the variational principle must be applied, are *defined* by

$$\mathbf{Q} \equiv \left(\xi \otimes \mathbf{1}\right)\mathbf{x}, \tag{5.18}$$

where

$$\xi \equiv \frac{1}{\sqrt{2}}\begin{pmatrix} -\mathbf{1} & \mathbf{1} \\ \mathbf{1} & \mathbf{1} \end{pmatrix} = \xi^{-1}. \tag{5.19}$$

Since $\xi\rho = \delta$, we recover the earlier solutions (with a trivial redefinition of the $y_{IA}$).

## VI. ELIMINATION OF THE BACKGROUND FRAME: SPONTANEOUSLY BROKEN LEPTON-HADRON SYMMETRY

Once again introducing the infinitesimal parameter $\varepsilon$, we replace the harmonic oscillator potential of the last section by

$$-\frac{1}{2N}\sum_{I=1}^{N}\sum_{J=1}^{N}\sum_{A=1}^{4}\sum_{B=1}^{4}\left(\mathbf{h}_4{}^2\right)_{AB}\left[x_{IA}(s)-x_{JB}(s)\right]^2, \tag{6.1}$$

where

$$\mathbf{h}_4 \equiv \left(1+2i\varepsilon\right)^{1/2}\mathbf{g}_4 + \varepsilon\mathbf{d}_4, \tag{6.2}$$

and $\mathbf{d}_4$ is defined as

$$\mathbf{d}_4 \equiv \frac{1}{2}\begin{pmatrix}\overline{\mathbf{g}}_2 & \overline{\mathbf{g}}_2\\ \overline{\mathbf{g}}_2 & \overline{\mathbf{g}}_2\end{pmatrix} = \frac{1}{4}\begin{pmatrix}1&1&1&1\\1&1&1&1\\1&1&1&1\\1&1&1&1\end{pmatrix} = \mathbf{d}_4{}^2. \tag{6.3}$$

Since $\mathbf{g}_4\mathbf{d}_4 = \mathbf{d}_4\mathbf{g}_4 = 0$, we have

$$\mathbf{h}_4{}^2 = \left(1+2i\varepsilon\right)\mathbf{g}_4 + \varepsilon^2\mathbf{d}_4. \tag{6.4}$$

The matrix $\mathbf{h}_4{}^2$ can be diagonalized by the matrix $\rho$, but not by $\delta$. Rather than solve for the coordinates $x_{IA}$, however, we make a transformation to the physical quark coordinates $Q_{IA}$ ,which we define as

$$\mathbf{Q} \equiv \left(\xi\otimes\mathbf{1}\right)\mathbf{x}, \tag{6.5}$$

with $\xi$ given in the last section. Note that $\xi\rho = \delta$ and $\xi^{-1}\mathbf{g}_4\xi = \mathbf{g}_4$.

The potential term can now be written as $(\mathbf{HQ})^2$, where

$$\mathbf{H} \equiv \left(\mathbf{g}_4 - i\varepsilon_N\mathbf{f}_4\right)\otimes\mathbf{N} + i\varepsilon_N\,\mathbf{1}\otimes\mathbf{1}, \tag{6.6}$$

and $\mathbf{f}_4$ is

$$\mathbf{f}_4 \equiv \xi^{-1}\mathbf{d}_4\xi = \begin{pmatrix}0 & 0\\ 0 & \overline{\mathbf{g}}_2\end{pmatrix} = \mathbf{f}_4{}^2. \tag{6.7}$$

Note that $\mathbf{f}_4\ \mathbf{g}_4 = \mathbf{g}_4\mathbf{f}_4 = 0$. The new Lagrangian becomes

$$L = -\,m\omega\left[-(\mathbf{HQ})^2\,\dot{\mathbf{Q}}^2 + (\dot{\mathbf{Q}}^T\cdot\mathbf{HQ})^2\right]^{1/2}. \qquad (6.8)$$

To construct a Hamiltonian, we turn again to Dirac's constraint formulation. Two primary constraints result from L:

$$\Phi_1 \equiv \mathbf{P}^2 + m^2\omega^2(\mathbf{HQ})^2 \approx 0,$$
$$\Phi_2 \equiv \mathbf{P}^T\cdot\mathbf{HQ} \approx 0. \qquad (6.9)$$

It is important that, to first order as $\varepsilon \to 0$, there are only two secondary constraints:

$$\Phi_3 \equiv \mathbf{P}^T\cdot\mathbf{GP} - m^2\omega^2\mathbf{Q}^T\cdot\mathbf{GQ},$$
$$\Phi_4 \equiv \mathbf{P}^T\cdot\mathbf{GQ} \approx 0. \qquad (6.10)$$

Thus, the algebra of the Poisson brackets of the constraint functions is closed.

The Dirac Hamiltonian is given by $H = \sum_{i=1}^{4} v_i\Phi_i$. Further examination of the constraint procedure shows that a solution can be obtained by setting $v_2=v_3=v_4=0$. We take the Dirac Hamiltonian to be

$$H = (1/2m)\left[\mathbf{P}^2 + m^2\omega^2(\mathbf{HQ})^2\right]. \qquad (6.11)$$

Define a transformation of coordinates

$$Q_{IA} = y_{IA} + N^{-1/2}\{\delta\,\mathbf{W}\}_A, \qquad \sum_{I=1}^{N} y_{IA} = 0. \qquad (6.12)$$

The momentum conjugate to $Q_{IA}$ is

$$P_{IA} = m\dot{Q}_{IA} = P_{IA} = p_{IA} + N^{-1/2}\{\delta\,\mathbf{p}\}_A, \qquad (6.13)$$

where the momenta $p_{IA}$ and $p_A$ are

$$p_{IA} \equiv m\,\dot{y}_{IA}, \qquad \sum_{I=1}^{N} p_{IA} = 0; \qquad p_A \equiv m\dot{W}_A. \qquad (6.14)$$

In terms of the transformed coordinates, the Hamiltonian becomes

$$H = \frac{1}{2m}\left\{\sum_{I=1}^{N}\sum_{A=1}^{4}\left[p_{IA}{}^2 - (\varepsilon m\omega)^2 y_{IA}{}^2\right] + \left[p_1{}^2 - (\varepsilon m\omega)^2 W_1{}^2\right] + p_3{}^2 + \sum_{A=2,4}\left[p_A{}^2 + (1+i\varepsilon)^2(m\omega)^2 W_A{}^2\right]\right\}. \qquad (6.15)$$

Thus, as in Sec. IV, the Hamiltonian still takes the harmonic-oscillator form but includes "repulsive" harmonic-oscillator potentials. $W_3$ is the only variable which is linear in s. Express the remaining solutions as

$$y_{IA} = (\varepsilon m\omega)^{-1}\left[a_{IA}\exp(\varepsilon\omega s) + b_{IA}\exp(-\,\varepsilon\omega s)\right],$$

$$p_{IA} = \left[ a_{IA} \exp(\varepsilon\omega s) - b_{IA} \exp(-\varepsilon\omega s) \right];$$

$$W_1 = \sqrt{2N}\,(\varepsilon m\omega)^{-1} \left[ a_1 \exp(\varepsilon\omega s) + b_1 \exp(-\varepsilon\omega s) \right],$$

$$p_1 = \sqrt{2N} \left[ a_1 \exp(\varepsilon\omega s) - b_1 \exp(-\varepsilon\omega s) \right];$$

$$W_A = \left( \frac{N\hbar}{m\omega} \right)^{1/2} \left[ a_A^{\dagger} \exp\left(i(1+i\varepsilon)\omega s\right) + a_A \exp\left(-i(1+i\varepsilon)\omega s\right) \right],$$

$$p_A = i\,(N\hbar m\omega)^{1/2}\,(1+i\varepsilon) \left[ a_A^{\dagger} \exp\left(i(1+i\varepsilon)\omega s\right) - a_A \exp\left(-i(1+i\varepsilon)\omega s\right) \right], A = 2,4. \quad (6.16)$$

Imposing the quantum conditions

$$\left[ y_{IA\mu}, p_{IA\nu} \right] = -i\hbar\, g_{\mu\nu},\ I = 1,2,3,\ldots,N-1; \qquad \left[ W_{A\mu}, p_{A\nu} \right] = -i\hbar\, 2N g_{\mu\nu}, \quad (6.17)$$

we have, to first order

$$\left[ a_{IA\mu}, b_{IA\nu} \right] = 0; \quad \left[ a_{1\mu}, b_{1\nu} \right] = 0; \quad \left[ a_{A\mu}, a_{A\nu}^{\dagger} \right] = -g_{\mu\nu},\ A = 2,4. \quad (6.18)$$

The variables $W_3$ and $p_3$ are proportional to the total center-of-mass and the total momentum, respectably, in the original system of coordinates $x_{IA}$. From the above solutions, we see that composite particles can be formed only in the asymptotic regions of s where $W_3$ and $p_3$ play no role. However, the constant $p_3$ appears in the constraint $H \approx 0$. We have

$$p_3 = m\dot{W}_3 = (2N)^{-1/2} \sum_{I=1}^{N} \sum_{A=1}^{4} m\dot{x}_{IA}. \quad (6.19)$$

Assuming the total momentum of the universe is finite or zero and that N is large, we shall neglect the term $p_3^{\,2}$ in the Hamiltonian. We have, then, to first order,

$$H = \frac{-2}{m} \left[ \sum_{I=1}^{N} \sum_{A=1}^{4} a_{IA} \cdot b_{IA} + 2N\, a_1 \cdot b_1 + N m^2 c^2 \sum_{A=2,4} \left( n_A + 2 \right) \right]. \quad (6.20)$$

The solutions for $W_A$ and $p_A$, A=2, 4, are complex numbers. As in Sec. IV, we have $W_A + W_A^{\dagger} \simeq W_A$ and $p_A + p_A^{\dagger} \simeq p_A$ as $s \to \pm\infty$ and $\varepsilon \to 0$. Again we shall retain the notation $W_A$ and $p_A$ with the understanding that we mean $W_A+W_A^{\dagger}$ and $p_A+p_A^{\dagger}$ .

To first order, commuting operators are $a_{IA}$, $b_{IA}$, $a_1$, $b_1$, and the number operators $n_A \equiv -a_A^{\dagger} a_A$. Write the state vectors as

$$|\Psi\rangle \equiv \left| a_{IA}, b_{IA}, a_1, b_1, n_2, n_4 \right\rangle. \quad (6.21)$$

The natural boundary conditions are $P_{IA}^{\ 2}(s) \sim 0$ as $s \to \pm\infty$, and yield

Hadron quarks (A=1, 2)

$$\left(a_{IA} + a_1\right)^2 |\Psi\rangle = \left(b_{IA} + b_1\right)^2 |\Psi\rangle = m^2c^2\left(n_2 + 2\right)|\Psi\rangle, \qquad (6.22)$$

$$a_2^{\ 2}|\Psi\rangle = 0, \quad \left(a_{IA} + a_1\right)\cdot a_2|\Psi\rangle = 0, \quad \left(b_{IA} + b_1\right)\cdot a_2|\Psi\rangle = 0. \qquad (6.23)$$

Lepton quarks (A=3, 4)

$$a_{IA}^{\ 2}|\Psi\rangle = b_{IA}^{\ 2}|\Psi\rangle = m^2c^2\left(n_4 + 2\right)|\Psi\rangle, \qquad (6.24)$$

$$a_4^{\ 2}|\Psi\rangle = 0, \quad a_{IA}\cdot a_4|\Psi\rangle = 0, \quad b_{IA}\cdot a_4|\Psi\rangle = 0. \qquad (6.25)$$

The n.b.c. lead to the satisfaction of the Dirac constraints other than $H \approx 0$.

There is one more condition that is appropriate to add for initial-state composite particles. An argument by Feynman[28] demonstrates why quantum mechanics and the relativistic restriction of particle velocities less than c require the existence of antiparticles. Although the quark velocities of this model are not so restricted, the velocities of the observable composites are. To produce composite particles which are accompanied by corresponding antiparticles, we shall introduce a definition of an antiquark. Recall that an electron going backward in time can be re-interpreted as a positron of opposite spin going forward in time. Let us also assume that the sign of $W_1$ is reversed and that an initial-state observable hadron is composed of a quark and an antiquark or

$$a_1 = -b_1. \qquad (6.26)$$

## VII. CLUSTER DECOMPOSITION:
## QUARK CONFINEMENT AND COMPOSITE-PARTICLE SCATTERING

As discussed in Sec. IV, the Lagrangian allows the straightforward consideration of cluster decomposition, which is necessary to describe the kinematically uncorrelated results of distant experiments. Note the quark coordinates are independent of N. Consider, for example, a cluster of four hadron quarks corresponding to I =1, 2. Kinematic decoupling is created by setting

$$\sum_{I=3}^{N} a_{IA} = \sum_{I=3}^{N} b_{IA} = 0; \quad \sum_{I=3}^{N}\sum_{A} a_{IA}\cdot b_{IA} = 0,\ A = 1,2; \quad \sum_{I=1}^{N}\sum_{A} a_{IA}\cdot b_{IA} = 0,\ A = 3,4. \qquad (7.1)$$

It follows that

$$\sum_{I=1}^{2} a_{IA} = \sum_{I=1}^{2} b_{IA} = 0, \quad A = 1,2. \qquad (7.2)$$

Since the quarks outside the subgroup form a cluster as well, we have

$$\sum_{I=1}^{2} \sum_{A=1,2} a_{IA} \cdot b_{IA} = 0. \qquad (7.3)$$

We will limit the discussion in this paper to clusters of four hadron quarks or four lepton quarks.

Denote eigenstates for the hadron cluster as $|\Psi\rangle \equiv |a_{IA}, b_{IA}, a_1, b_1, n_2, n_4\rangle$, I=1, 2, and A=1, 2.

Note that, $n_2$ and $n_4$ can take on positive or negative values. We shall consider initial states of real mass that satisfy the following:

(1) Cluster decomposition conditions $\sum_{I=1}^{2} a_{IA} = \sum_{I=1}^{2} b_{IA} = 0$, and $\sum_{I=1}^{2} \sum_{A=1,2} a_{IA} \cdot b_{IA} = 0$.

(2) The quark-antiquark assumption $a_1 = -b_1$.

(3) The n.b.c. equations.

(4) The constraint $H \approx 0$, or, to first order,

$$\left[ a_1 \cdot b_1 + \left(m^2c^2/2\right) \sum_{A=2,4} (n_A + 2) \right] |\Psi\rangle = 0. \qquad (7.4)$$

For initial lepton cluster states, we consider a similar set of requirements.

## A. Lepton-lepton scattering

For the decoupled subgroup of four lepton quarks, the quark position vectors are

$$Q_{IA} = \frac{1}{\varepsilon m\omega}\left[ a_{IA} \exp(\varepsilon\omega s) + b_{IA} \exp(-\varepsilon\omega s)\right] + \frac{1}{\sqrt{2N}} W_3$$
$$+ (-1)^{A-3} \sqrt{2}\, l \cosh(\varepsilon\omega s) \left[ a_4^{\dagger} \exp(i\omega s) + a_4 \exp(-i\omega s)\right], \qquad (7.5)$$

where $I = 1,2;\ A = 3,4$. From cluster decomposition, it follows that,

$$a_{13} = -a_{23},\ a_{14} = -a_{24}, b_{13} = -b_{23},\ b_{14} = -b_{24},\ a_{13} \cdot b_{13} = -a_{14} \cdot b_{14}. \qquad (7.6)$$

From the discussion of Sec. IV it is straightforward to find that lepton-lepton scattering, regardless of energy, takes place only in the forward or backward directions. Since the leptons are identical, one cannot discern that any scattering has taken place. In other words, the leptons behave as if they are point particles.

## B. Hadron-hadron scattering

Now consider a decoupled cluster of four hadron quarks:

$$Q_{IA} = (1/\varepsilon m\omega)\Big[\big(a_{IA} + a_1\big)\exp(\varepsilon\omega s) + \big(b_{IA} + b_1\big)\exp(-\varepsilon\omega s)\Big]$$

$$+ (-1)^{A-1}\sqrt{2}\, l \cosh(\varepsilon\omega s)\Big[a_2^{\dagger}\exp(i\omega s) + a_2\exp(-i\omega s)\Big]\Big\}, \quad (7.7)$$

where $I = 1,2;\;\; A = 1,2$ . The kinematic decoupling from the remainder of the system implies

$$\sum_{I=1}^{2} a_{IA} = \sum_{I=1}^{2} b_{IA} = 0, \qquad A = 1,2; \quad \text{and} \quad \sum_{I=1}^{2}\sum_{A=1,2} a_{IA}\cdot b_{IA} = 0. \quad (7.8)$$

Looking at a particular case, we include in the initial conditions, at $t = -\infty$ the following quark confinement relation:

$$a_{11} = a_{22} \;\; \text{where } a_{110} < 0. \quad (7.9)$$

As in the lepton four-quark cluster, quark confinement in three other hadrons results from cluster decomposition. The following hadrons exist at $s = \pm\infty$:

$$Q_1^{+H} \equiv \tfrac{1}{2}(Q_{11} + Q_{22})\Big|_{s\to+\infty} \sim (1/\varepsilon m\omega)\big(a_{11} + a_1\big)\exp(\varepsilon\omega s),$$

$$Q_2^{+H} \equiv \tfrac{1}{2}(Q_{21} + Q_{12})\Big|_{s\to+\infty} \sim (1/\varepsilon m\omega)\big(-a_{11} + a_1\big)\exp(\varepsilon\omega s),$$

$$Q_1^{-H} \equiv \tfrac{1}{2}(Q_{11} + Q_{12})\Big|_{s\to-\infty} \sim (1/\varepsilon m\omega)\big(b_{11} + b_1\big)\exp(-\varepsilon\omega s),$$

$$Q_2^{-H} \equiv \tfrac{1}{2}(Q_{21} + Q_{22})\Big|_{s\to-\infty} \sim (1/\varepsilon m\omega)\big(-b_{11} + b_1\big)\exp(-\varepsilon\omega s). \quad (7.10)$$

The hadron internal states are described by

$$q^H(s) \equiv \sqrt{2}\, l \exp(|\varepsilon\omega s|)\Big[a_2^{\dagger}\exp(i\omega s) + a_2\exp(-i\omega s)\Big], \quad (7.11)$$

yielding, as $\varepsilon \to 0$,

$$|q^H| = 2l\sqrt{n_2 + 2}\,. \quad (7.12)$$

The hadron momenta are

$$P_1^{+H} \equiv \big(P_{11} + P_{22}\big)\Big|_{s\to+\infty} \sim 2\big(a_{11} + a_1\big)\exp(\varepsilon\,\omega s),$$

$$P_2^{+H} \equiv \big(P_{21} + P_{12}\big)\Big|_{s\to+\infty} \sim 2\big(-a_{11} + a_1\big)\exp(\varepsilon\,\omega s),$$

$$P_1^{-H} \equiv \big(P_{11} + P_{12}\big)\Big|_{s\to-\infty} \sim -2\big(b_{11} + b_1\big)\exp(-\varepsilon\,\omega s),$$

$$P_2^{-H} \equiv \big(P_{21} + P_{22}\big)\Big|_{s\to-\infty} \sim -2\big(-b_{11} + b_1\big)\exp(-\varepsilon\,\omega s). \quad (7.13)$$

From the n.b.c., we obtain the mass-shell constraints

$$\left(P_I^{\pm H}\right)^2 = m_{n_2}{}^2 c^2 = 4m^2c^2\left(n_2 + 2\right). \tag{7.14}$$

Total four-momentum is conserved:

$$P_1^{+H} + P_2^{+H} + P_1^{-H} + P_2^{-H} = 0. \tag{7.15}$$

As noted for lepton scattering, the cluster space-time emits and absorbs free composite particles in an observer's frame, but is otherwise not observable.

In the particular case above, if we further stipulate $b_{110} < 0$, two quarks, $Q_{11}$ and $Q_{21}$, turn around in time. A different configuration can be chosen such that quarks $Q_{12}$ and $Q_{22}$ do the same. The two configurations are similar in kind, but differ in the assignment of quarks in the initial state. Besides these configurations, a third type is possible where none of the four quarks turn around in time (this is not an option in lepton-lepton scattering).

The following picture of hadron-hadron scattering arises. When $s \to \pm\infty$, free composite particles, namely hadrons, are formed in the asymptotic regions of the cluster space-time. We take the initial (final) state of the particles in the observer's frame to be the initial (final) state in the cluster space-time. The scattering is similar to lepton-lepton scattering in the following ways: (1) Scattering takes place by quark exchange; (2) quark-antiquark confinement in the initial state ($t = -\infty$) implies quark-antiquark confinement in the final state ($t = +\infty$); (3) total four-momentum is conserved; (4) the hadrons all have zero orbital angular momentum, and equal internal angular momentum. However, hadron scattering can occur at angles other than $0^o$ and $180^o$. We will see that the internal variable $a_1$ plays the role of an impact parameter (recall the absence of the comparable term for leptons). Just as for lepton scattering, the model predicts that hadron quarks are observable only indirectly as constituents of composite particles.

## C. Hadron-hadron scattering amplitudes

In the usual formulation of scattering amplitudes, one employs two complete sets of free-particle state vectors to represent the "incoming" and "outgoing" states of the system. One goes from the "IN" basis to the "OUT" basis by means of a unitary transformation, or S-matrix. In the present case, however, the S-matrix formulation does not apply. First, there is no evolution of the state vector, since H vanishes. Second, the state vector contains the eigenvalues of all the operators corresponding to $t = \pm\infty$. Therefore, it is meaningless to speak of the transition of the state describing the hadrons at $t = -\infty$ to a state at $t = +\infty$.

Physical measurements at $t = -\infty$ do not determine all of the state eigenvalues. Furthermore, the incoming hadron momenta and masses do not determine which of the four quarks are involved. Thus the state vector must be expressed as a linear combination of physical states corresponding to all possible configurations that might arise. We shall assume, for simplicity, that these states have equal probability of occurring.

Denote the incoming hadron momenta as $P_1$ and $P_2$. The three possible quark configurations are defined as the s-, t-, and u-channel configurations, respectively. For the s-channel, put

$$P_1 \equiv P_1^{-H} = -2\left(b_{11} + a_1\right),$$

$$P_2 \equiv P_2^{-H} = -2\left(-b_{11} + a_1\right). \tag{7.16}$$

The state vector at $t = -\infty$ is

$$\left|\Psi_{IN}^{s-ch}\right\rangle = \sum_{n_3=-\infty}^{\infty} \prod_{I=1}^{2} \prod_{A=1,2} \int d^4a_{IA} \int d^4b_{IA} \int d^4a_1\, \delta\left(\frac{2}{(mc)^2}\, a_1^{\,2} - \left(n_2 + n_4 + 4\right)\right)$$

$$\times\, \delta\left(P_1 + 2\left(b_{11} + a_1\right)\right)\, \delta\left(P_2 - 2\left(b_{11} - a_1\right)\right)\left|a_{IA}, b_{IA}, a_1, b_1 = -a_1, n_2, n_4\right\rangle. \tag{7.17}$$

The initial states are further taken to satisfy the n.b.c. and cluster decomposition.

The physical initial conditions imply that the final state consists also of two composites, each of mass $m_{n_4}$. Label the composite momenta at $t = +\infty$ as $P_3$ and $P_4$. Then, it follows that

$$P_3 \equiv P_1^{+H} = 2\left(a_{11} + a_1\right),$$
$$P_4 \equiv P_2^{+H} = 2\left(a_{21} + a_1\right). \tag{7.18}$$

Express the outgoing state in the s-channel as

$$\left\langle\Psi_{OUT}^{s-ch}\right| = \sum_{n_3=-\infty}^{\infty} \prod_{I=1}^{2} \prod_{A=1,2} \int d^4a_{IA} \int d^4b_{IA} \int d^4a_1 \int d^4b_1$$

$$\times\, \delta\left(P_3 - 2\left(a_{11} + a_1\right)\right)\, \delta\left(P_4 - 2\left(a_{21} + a_1\right)\right)\left\langle a_{IA}, b_{IA}, a_1, b_1, n_2, n_4\right|. \tag{7.19}$$

Calculating the overlap of these two state vectors yields

$$\left\langle\Psi_{OUT}^{s-ch}\middle|\,\Psi_{IN}^{s-ch}\right\rangle \propto \delta\left(P_1 + P_2 + P_3 + P_4\right)\prod_{I=1}^{4}\delta\left(P_I^{\,2} - m_{n_2}^{\;\;2}c^2\right)\sum_{n=-\infty}^{\infty}\delta\left(\frac{\left(P_1 + P_2\right)^2}{8m^2c^2} - n\right). \tag{7.20}$$

The t-channel and u-channel contributions are treated similarly. The state vectors for the three channels are orthogonal, and the elastic hadron-hadron scattering amplitude becomes

$$A(s,t,u) = N\left[ \left\langle \Psi^{s-ch}_{OUT} \middle| \Psi^{s-ch}_{IN} \right\rangle + \left\langle \Psi^{t-ch}_{OUT} \middle| \Psi^{t-ch}_{IN} \right\rangle + \left\langle \Psi^{u-ch}_{OUT} \middle| \Psi^{u-ch}_{IN} \right\rangle \right]$$

$$= N \prod_{I=1}^{4} \delta\left(P_I^{\,2} - m_{n_2}^{\,2} c^2\right) \delta\left(P_1 + P_2 + P_3 + P_4\right)\left[D(s) + D(t) + D(u)\right], \tag{7.21}$$

where

$$D(z) \equiv \sum_{n=-\infty}^{\infty} \delta(\alpha z - n), \qquad \alpha \equiv \left(8m^2c^2\right)^{-1} \equiv \left(m_R c\right)^{-2}, \tag{7.22}$$

and s, t, and u are the Mandelstam variables

$$s = \left(P_1 + P_2\right)^2; \quad t = \left(P_1 + P_3\right)^2; \quad u = \left(P_1 + P_4\right)^2. \tag{7.23}$$

and $N$ is a proportionality constant. In the center-of-energy system of the incoming composites,

$$s = (E_{CE}/c)^2,$$

$$t = -2q^2(1 + \cos\Theta),$$

$$u = -2q^2(1 + \cos\Theta), \tag{7.24}$$

where q and $\Theta$ are the three-momentum and scattering angle, respectively. The variable s here should not be confused with the evolution parameter.

The contributing values of s, t, and u to the invariant scattering amplitude, i.e. the poles of the three terms, satisfy

$$\alpha s + \alpha t + \alpha u = 1. \tag{7.25}$$

The first maximum in the physical s-channel occurs for $s = 1/\alpha, t = 0, u = 0$. It corresponds to the lowest-mass physical composite particle. In fact, rewriting the composite mass squared as

$$m_{n_2}^{\,2} = m_R^{\,2}\left(1 + \frac{n_2}{2}\right), \tag{7.26}$$

we see, that $s = n/\alpha = n\left(m_R c\right)^2$, $n = 1, 2, 3, 4, \ldots$, corresponds to the composite masses $m_0$, $m_2$, $m_4$, $m_6$, ... , respectively. We shall see in the next section that these particles have angular momenta 0, 2, 4, 6, ... respectively, or $s = n\left(m_R c\right)^2$ corresponds to resonances of angular momenta $l$ =2(n-1).

The amplitudes arise from a superposition of states corresponding to the three quark scattering diagrams. Resonance production in the s-channel is suggested by maxima in s and for a given s, the angular distribution has maxima as well. Can we make a connection to the internal angular momenta of these implied resonances? The internal angular momentum of composite particles is examined in the next section where it is shown that the angular momentum quantum number $l$ equals the mass quantum number $n_A$. Each scattering amplitude above, for a given "resonance" $s = n\left(m_R c\right)^2$, has a dependance on the scattering angle $\Theta$, with maxima at

$$t = 0, -1/\alpha, -2/\alpha, \ldots, -(n-1)/\alpha \qquad (u = -(n-1)/\alpha, \ldots, -1/\alpha, 0). \tag{7.27}$$

The plot of the angular dependence of the scattering amplitude when $s = n/\alpha$ has an interesting similarity to the function,

$$1 + 2P_{2(n-1)}(\cos\Theta), \tag{7.28}$$

where $P_{2(n-1)}$ are the Legendre functions. This again suggests that each resonance is associated with a particular angular momentum $l = 2(n-1)$.

Let us now compare these scattering amplitudes to amplitudes for the simplest dual model, the Veneziano model.[23] The scattering amplitude for the elastic scattering of two identical (two-quark) mesons is given by the ad hoc assumption

$$A(s,t,u) = g^2\left[V(s,t) + V(t,u) + V(u,s)\right],$$
$$V(s,t) = \frac{\Gamma(-\alpha(s))\,\Gamma(-\alpha(t))}{\Gamma(-\alpha(s)-\alpha(t))} \equiv B(-\alpha(s), -\alpha(t)), \tag{7.29}$$

where $\alpha(z) = \alpha_0 + \alpha' z$; $g, \alpha_0$, and $\alpha'$ are constants, and $\Gamma$ and B are Euler functions. The function $V(s,t)$ has simple poles at $\alpha(s) = 0, 1, 2, \ldots$, with residues which depend polynomially on t. A further condition is usually assumed, such as

$$\alpha(s) + \alpha(t) + \alpha(u) = -1. \tag{7.30}$$

In the Veneziano amplitude, there are no singularities of V(s,t) except for poles on the real semi-axis, and the residue at each pole in one of the variables is a polynomial in the other variable. The poles are interpreted as particle resonances. The residue at the *k*-th pole in the variable s is a polynomial in t of degree *k*. This implies a degeneracy in the masses of the resonances with different angular momenta. In other words, the residue does not reduce to $P_k(\cos\Theta)$, which would indicate a resonance of angular momentum *l*=*k*. It should be remembered that spin is not included in either model.

Finally, it is interesting that while hadron-hadron scattering produces "resonances," lepton-lepton scattering does not. Consider a universe described by a model of this type which describes both real hadrons and real leptons in the same solution (the leptons need not be the same mass as the hadrons). This can be obtained, for example, by adding another identical system of quarks to the Lagrangian of Sec. VI (as we did in Sec. V). An observer sees hadrons of a given mass scatter from each other giving rise to excited states (resonances) which immediately decay back into the hadrons of the given masses again. No excited states of the leptons occur.

## VIII. INTERNAL ANGULAR MOMENTUM

Although spin has been omitted, the composites have an internal angular momentum, which if specified in the initial conditions will modify the Veneziano-type scattering amplitudes. Although such amplitudes are not considered in this article, it was mentioned in the last section that if the hadron amplitudes are related to resonance production, the resonance angular momentum number equals its mass number $n_2$. That is, there is no mass degeneracy for a given angular momentum. We shall demonstrate that in this section.

First, we discuss the complete specification of the physical states. For simplicity, the discussion is limited to leptons. The generalization to hadron states will be obvious.

### A. Compete set of number states

The specification of a momentum lepton-quark state in terms of the number eigenvalues is given by $\left|p_\mu, n_1, n_2, n_3, n_0\right\rangle$, where the momenta of the other quarks in the system are understood to be implicitly included, and

$$n_i \equiv a_i^\dagger \cdot a_i, \quad i = 1,2,3; \qquad n_0 \equiv a_0^\dagger \cdot a_0, \tag{8.1}$$

and, we recall, $n \equiv -a^\dagger \cdot a$.

The number representation for the spatial components is well known. For the one-dimensional case, operators a and $a^\dagger$ obey the commutation relation

$$\left[a, a^\dagger\right] = 1. \tag{8.2}$$

Defining $n = a^\dagger a$, we find

$$\left[a, n\right] = a; \qquad \left[a^\dagger, n\right] = -a^\dagger. \tag{8.3}$$

This implies that $a|n\rangle$ and $a^\dagger|n\rangle$ are eigenstates of n with eigenvalues $n-1$ and $n+1$, respectively. Assume $\langle n|\, n'\rangle = \delta_{n,n'}$ and also

$$\langle n|a^\dagger a|n\rangle \geq 0. \tag{8.4}$$

Then n is also positive or zero. The lowest value zero implies

$$a|0\rangle = 0. \tag{8.5}$$

while $a^\dagger|0\rangle$ is an eigenstate of n with eigenvalue 1. In the same manner, one finds that the values of n are 0, 1, 2, 3, ... . The normalization is taken to be

$$a|n\rangle = \sqrt{n}\,|n-1\rangle; \quad a^\dagger|n\rangle = \sqrt{n+1}\,|n+1\rangle. \tag{8.6}$$

To describe the analogous representation for operators $a_\mu$ and $a_\mu^\dagger$, it is sufficient to consider two dimensions, the z-axis and the time axis. The commutation relations are

$$\left[a_3, a_3^\dagger\right] = 1; \qquad \left[a_0, a_0^\dagger\right] = -1. \tag{8.7}$$

Defining $n = n_3 - n_0$, we have

$$\left[a_3, n\right] = a_3, \ \left[a_3^{\dagger}, n\right] = -a_3^{\dagger}, \ \left[a_0, n\right] = -a_0, \ \left[a_0^{\dagger}, n\right] = a_0^{\dagger}. \tag{8.8}$$

From these relations, it follows that for the state $a_3 \left|n_3, n_0\right\rangle$, $n_3$ has the eigenvalue $n_3$-1, while for the state $a_0 \left|n_3, n_0\right\rangle$, $n_0$ has the eigenvalue $n_0$+1. Assume for real mass composites

$$n = \left\langle n_3, n_0\right| \left(a_3^{\dagger} \cdot a_3 - a_0^{\dagger} \cdot a_0\right) \left|n_3, n_0\right\rangle = \left\langle n_3, n_0\right| \left(n_3 - n_0\right) \left|n_3, n_0\right\rangle \geq 0. \tag{8.9}$$

Write

$$a_0 \left|n_0\right\rangle = N_{n_0} \left|n_0 + 1\right\rangle. \tag{8.10}$$

Then

$$n_0 \left\langle n_0 \middle| n_0\right\rangle = \left\langle n_0\right| a_0^{\dagger} \ a_0 \left|n_0\right\rangle = N_{n_0}^{\ *} N_{n_0} \left\langle n_0 + 1 \middle| n_0 + 1\right\rangle. \tag{8.11}$$

We put

$$\left\langle n_0 \middle| n_0{}'\right\rangle = (-1)^{n_0} \delta_{n_0, n_0'}. \tag{8.12}$$

Therefore, $n_0$ takes on the values 0, -1, -2, -3, ..., and thus has an upper bound of zero.

We can now write

$$\left\langle n_1, n_2, n_3, n_0 \middle| n_1, n_2, n_3, n_0\right\rangle = (-1)^{n_0}. \tag{8.13}$$

The reader should not be alarmed at the negative probabilities, as we have not yet expressed the physical states in terms of the number states. There are no negative probabilities for the physical states.

Choose the following normalization constants:

$$a_0 \left|n_0\right\rangle = \exp\left[-i\frac{\pi}{2}(n_0 + 1)\right] \sqrt{-n_0} \left|n_0 + 1\right\rangle,$$

$$a_0^{\dagger} \left|n_0\right\rangle = \exp\left[i\frac{\pi}{2} n_0\right] \sqrt{-n_0 + 1} \left|n_0 - 1\right\rangle. \tag{8.14}$$

Thus, for example,

$$a_0 \left|0\right\rangle = 0, \quad a_0 \left|-1\right\rangle = \left|0\right\rangle, \quad a_0 \left|-2\right\rangle = i\sqrt{2} \left|-1\right\rangle,$$
$$a_0^{\dagger} \left|0\right\rangle = \left|-1\right\rangle, \quad a_0^{\dagger} \left|-1\right\rangle = i\sqrt{2} \left|-2\right\rangle. \tag{8.15}$$

### B. Physical states: A subset of the number states

The set of states that satisfy the n.b.c. are linear combinations of a subset of the number states $\left|p_{\mu}, n_1, n_2, n_3, n_0\right\rangle$. Begin with the subset of states which satisfy the one eigenvalue equation of the n.b.c., namely, the mass-shell relation, and then find the linear combinations of these states that satisfy the remaining constraints $a^2 = 0$, and $p \cdot a = 0$. The discussion is limited to n = 0, 1, and 2.

Define

$$\beta \equiv \left(1 - \frac{\mathbf{p}^2}{p_0^{\;2}}\right)^{-1/2}. \tag{8.16}$$

If the momentum is along the z-axis, the following linear combinations satisfy the n.b.c.

n=0: $\quad |p_3,0,0,0,0\rangle;$

n=1: $\quad |p_3,1,0,0,0\rangle,\ \ |p_3,0,1,0,0\rangle,\ \ \beta\left[|p_3,0,0,1,0\rangle + \frac{p_3}{p_0}|p_3,0,0,0,-1\rangle\right];$

n=2: $\quad |p_3,1,1,0,0\rangle,$

$$\beta\left[|p_3,1,0,1,0\rangle + \frac{p_3}{p_0}|p_3,1,0,0,-1\rangle\right],\quad \beta\left[|p_3,0,1,1,0\rangle + \frac{p_3}{p_0}|p_3,0,1,0,-1\rangle\right],$$

$$\frac{1}{\sqrt{2}}\left\{|p_3,2,0,0,0\rangle - \beta^2\left[|p_3,0,0,2,0\rangle + \sqrt{2}\,\frac{p_3}{p_0}|p_3,0,0,1,-1\rangle - i\left(\frac{p_3}{p_0}\right)^2|p_3,0,0,0,-2\rangle\right]\right\},$$

$$\frac{1}{\sqrt{2}}\left\{|p_3,0,2,0,0\rangle - \beta^2\left[|p_3,0,0,2,0\rangle + \sqrt{2}\,\frac{p_3}{p_0}|p_3,0,0,1,-1\rangle - i\left(\frac{p_3}{p_0}\right)^2|p_3,0,0,0,-2\rangle\right]\right\}. \tag{8.17}$$

These states are normalized to unity, but they are not all orthogonal. Note that the conditions $a^2$=0 and $p\cdot a = 0$ reduces the number of independent states for n=2 from six to five states.

## C. Angular momentum states for real-mass leptons

The next step is to define linear combinations of the above states for a given value of n that are both normalized and orthogonal. Consider the real-mass lepton in its rest frame. The internal angular momentum is given by

$$\boldsymbol{l} = i\hbar\ \mathbf{a}^\dagger \times \mathbf{a}. \tag{8.18}$$

The conditions $a^2 = 0$ and $p\cdot a = 0$ lead to[11]

$$|\boldsymbol{l}|^2 = (\hbar)^2\ n(n+1). \tag{8.19}$$

In other words, the orbital angular momentum label $l$ is equal to n. (The inclusion of spin can produce a degeneracy in n.) As is well known, this produces particles on a Regge trajectory. Recall also that the *nonrelativistic* three-dimensional harmonic oscillator has, in general, more than one angular momentum for a given n.

Consider now the eigenstates of the commuting operators

$$p_\mu, n,\ |\boldsymbol{l}|^2, l_3 = i\hbar\ \left(a_1^{\;\dagger}a_2 - a_1 a_2^{\;\dagger}\right). \tag{8.20}$$

Since n = $l$, there are 2n+1 eigenstates of $l_3$ for a given n. Label rest frame states that are eigenstates of the mass-shell condition as $\left| n, l = n; \mathbf{p} = 0, l_3 \right\rangle$. In terms of the number states $\left| \mathbf{p} = 0, n_1, n_2, n_3, n_0 \right\rangle$, the rest frame states for n=0, 1, and 2 are the following:

n=0: $\left|0,0; \mathbf{p}=0, l_3=0\right\rangle = |\mathbf{p}=0,0,0,0,0\rangle$;

n=1: $\left|1,1; \mathbf{p}=0, l_3=\pm 1\right\rangle = \frac{1}{\sqrt{2}}\left[|\mathbf{p}=0,1,0,0,0\rangle \mp i|\mathbf{p}=0,0,1,0,0\rangle\right]$,

$\left|1,1; \mathbf{p}=0, l_3=0\right\rangle = |\mathbf{p}=0,0,0,1,0\rangle$;

n=2: $\left|2,2; \mathbf{p}=0, l_3=\pm 2\right\rangle = \frac{1}{2}\left[|\mathbf{p}=0,2,0,0,0\rangle - |\mathbf{p}=0,0,2,0,0\rangle\right] \mp \frac{1}{\sqrt{2}}|\mathbf{p}=0,1,1,0,0\rangle$,

$\left|2,2; \mathbf{p}=0, l_3=\pm 1\right\rangle = \frac{1}{\sqrt{2}}\left[|\mathbf{p}=0,1,0,1,0\rangle \mp |\mathbf{p}=0,0,1,1,0\rangle\right]$,

$$\left|2,2; \mathbf{p}=0, l_3=0\right\rangle = \frac{1}{\sqrt{6}}\left[|\mathbf{p}=0,2,0,0,0\rangle + |\mathbf{p}=0,0,2,0,0\rangle - 2|\mathbf{p}=0,0,0,2,0\rangle\right]. \quad (8.21)$$

These states are normalized to unity and are all orthogonal.

The next task is to express the above relations in an arbitrary reference frame. Note the similarity of the labels in $\left| n, l; \mathbf{p} = 0, l_3 \right\rangle$ to the labels for rest-frame single-particle states of the Poincaré algebra, namely $\left| M^2, s; \mathbf{p} = 0, s_3 \right\rangle$, where M is the mass and s the spin.

It is interesting to compare the operator algebra for the model to that for the Poincaré group. The Casimir operators for the Poincaré group are $P^2=M^2c^2$ and $W^2$, where $W_\mu$ is the Pauli-Lubansky operator

$$W_\mu = -\frac{1}{2}\varepsilon_{\mu\nu\lambda\rho}P^\nu J^{\lambda\rho}. \quad (8.22)$$

The generators of the Poincaré group, $P_\mu$ and $J_{\mu\nu}$, do not commute. Defining $J_{0i} \equiv K_i$, $J_{ij} \equiv \varepsilon_{ijk} J_k$, we can write

$$W_0 = \mathbf{P}\cdot\mathbf{J}, \qquad W_i = \varepsilon_{ijk} K_j P_k + J_i P_0. \quad (8.23)$$

Thus, in the rest frame, $W^2=-M^2c^2s(s+1)$. Labels $\mathbf{p}$ and $s_3$ come from the choice of diagonalization. For states of arbitrary momentum, it is customary to keep the spin label s and $s_3$, writing the state as $\left| M^2, s; \mathbf{p}, s_3 \right\rangle$.

Returning to the formulation in terms of the number operators $n_i$ and $n_0$, we define an operator similar to the Pauli-Lubansky operator, namely

$$w_\mu \equiv -\frac{1}{2}\,\varepsilon_{\mu\nu\lambda\rho}\,p^\nu\, l^{\lambda\rho}\,, \tag{8.24}$$

where we write

$$k_i \equiv l_{0i}\,,\text{and } l_i \equiv \varepsilon_{ijk}\, l^{jk}; \tag{8.25}$$

$$k_i = -i\hbar\;(a_0{}^\dagger a_i - a_0 a_i{}^\dagger), \qquad l_i = i\hbar\;\varepsilon_{ijk}\left(a_j{}^\dagger a_k - a_j a_k{}^\dagger\right). \tag{8.26}$$

We find that $k_i$ and $l_i$ obey the same algebra as $K_i$ and $J_i$. Of course, in contrast to the Poincaré group, $p_\mu$ and $\boldsymbol{l}$ commute. The components of $w_\mu$ are

$$w_0 = \boldsymbol{l}\cdot\mathbf{p}, \quad w_i = \varepsilon_{ijk} k_j p_k + l_i p_0\,. \tag{8.27}$$

Thus, in the rest frame of the particle, $w_0$=0 and $w^2 = -\,m_n{}^2 c^2\,|\boldsymbol{l}|^2$. In analogy to the Poincaré states, the basis states in an arbitrary frame will be labeled $\left|n, l = n;\, \mathbf{p}, l_3\right\rangle$.

As an example, consider a lepton whose three-momentum lies along the z-axis. The physical states must then obey then n.b.c. condition

$$p_3 a_3 - p_0 a_0 = 0\,. \tag{8.28}$$

The physical states in terms of the states $\left|\,p_3, n_1, n_2, n_3, n_0\right\rangle$ are given below:

n=0: $\quad \left|n = 0, l = 0;\, p_3, l_3 = 0\right\rangle = \left|p_3, 0,0,0,0\right\rangle;$

n=1: $\quad \left|n = 1, l = 1;\, p_3, l_3 = \pm 1\right\rangle = \dfrac{1}{\sqrt{2}}\left[\left|p_3, 1,0,0,0\right\rangle \mp i\left|p_3, 0,1,0,0\right\rangle\right],$

$$\left|n = 1, l = 1;\, p_3, l_3 = 0\right\rangle = \beta\left[\left|p_3, 0,0,1,0\right\rangle + \frac{p_3}{p_0}\left|p_3, 0,0,0,-1\right\rangle\right];$$

n=2: $\quad \left|\,n = 2, l = 2;\, p_3, l_3 = \pm 2\right\rangle = \dfrac{1}{2}\left[\left|p_3, 2,0,0,0\right\rangle - \left|p_3, 0,2,0,0\right\rangle\right] \mp \dfrac{i}{\sqrt{2}}\left|p_3, 1,1,0,0\right\rangle,$

$$\left|\,n = 2, l = 2;\, p_3, l_3 = \pm 1\right\rangle = \frac{1}{\sqrt{2}}\,\beta\,\Big\{\left[\left|p_3, 1,0,1,0\right\rangle \mp i\left|p_3, 0,1,1,0\right\rangle\right] + \frac{p_3}{p_0}\left[\left|p_3, 1,0,0,-1\right\rangle \mp i\left|p_3, 0,1,0,-1\right\rangle\right]\Big\},$$

$$\left|\,n = 2, l = 2;\, p_3, l_3 = 0\right\rangle = \frac{1}{\sqrt{6}}\Big\{\left[\left|p_3, 2,0,0,0\right\rangle + \left|p_3, 0,2,0,0\right\rangle\right]$$

$$-\,2\beta^2\left[\left|p_3, 0,0,2,0\right\rangle + \sqrt{2}\,\frac{p_3}{p_0}\left|p_3, 0,0,1,-1\right\rangle - i\left(\frac{p_3}{p_0}\right)^2\left|p_3, 0,0,0,-2\right\rangle\right]\Big\} \tag{8.29}$$

These states are normalized to unity. are orthogonal, and satisfy the n.b.c. for real mass particles.

Generalizing these results, we specify the state vectors for the hadron-hadron scattering example considered earlier as $\left|a_{IA}, b_{IA}, a_1, b_1, n_2, n_4; l = n_2, l_3\right\rangle$ where each hadron has internal angular momentum $l$. If we adopt the convention that, for hadrons going forward in time as s increases, $l_3$ is the projection of the hadron angular momentum along the z-axis, then for hadrons going backwards in time, the projection has the value $-l_3$. Thus, in general, if the initial internal angular momentum projection $l_3$ is specified as an initial condition, not all of the three scattering configurations discussed earlier contribute (see the derivation of the hadron scattering amplitude), and the Veneziano-type amplitude will be modified.

## IX. LARGE NUMBER HYPOTHESIS AND COSMIC IMPLICATIONS

Observed matter in the universe consists almost entirely of neutrons, protons, and electrons. Cosmologists define the $N_1$ group of numbers as dimensionless expressions that cluster around $10^{40}$ and involve nucleon or electron masses.[22] Examples are

$$N_1 \sim \frac{e^2}{Gm_{nuc}^{\ 2}} \sim \frac{hc}{Gm_{nuc}^{\ 2}} \sim 10^{40}, \tag{9.1}$$

where $m_{nuc}$ is the nucleon mass. A second group, the $N_2$ group, contains dimensionless numbers involving the expansion of the universe. These also cluster about $10^{40}$. For example,

$$N_2 \sim \frac{L_H}{l_{nuc}} \sim 10^{40}, \tag{9.2}$$

where $L_H$ is the Hubble length, i.e., the size of the observable universe, and $l_{nuc}$ is the nucleon size. The equation

$$N_1 \simeq N_2 \tag{9.3}$$

known as Dirac's large number hypothesis,[22,29] has been considered by many scientists to be too remarkable a coincidence not to be a permanent relationship. Yet an expanding universe implies $L_H$ is getting larger, while the quantities in $N_1$ are generally believed to be constant. For example, spectral lines from Star X, once redshifts are accounted for, are the same as here on Planet Earth indicating the atoms are the same both places.

In the present model, however, masses and nucleon sizes change as the universe expands. If we replace the nucleon mass $m_{nuc}$ in the $N_1$ expressions above by the ground-state mass

$$m_{n_2=0} \approx N_2^{\ -1/2} m_P, \tag{9.4}$$

we find $N_1=N_2$. Similarly, replacing $l_{nuc}$ in $N_2$ by

$$l_{n_4=0} \approx N_2^{\,1/2} l_P \,, \tag{9.5}$$

yields

$$N_2 \simeq \left(L_H/l_P\right)^{2/3}. \tag{9.6}$$

As $L_H$ grows larger, both $N_1$ and $N_2$ increase as well. For the observer on earth today, $N_2 \simeq 10^{40}$, and ground-state composite particles have the mass and size on the order of a nucleon.

The action-at-a-distance Lagrangian includes all 4N quarks in the universe, not just the observable universe. However, the choice of N does not affect the quark solutions; it affects only the cluster decomposition conditions. By imposing these conditions only on quarks in the observable universe, we see the same masses and sizes for all like particles on our backward light-cone (after accounting for the usual redshifts and special relativistic effects). However, an observer on earth at an earlier cosmic time would measure masses and sizes that are different multiples of the Planck constants.

Dirac, in questioning whether mass can change over cosmic time,[29] remarked we would then expect ancient rock crystals to change their crystalline form. He rejected the assumption of the continuous creation of matter. However, as we show in the next section, the simultaneous change in size and mass suggest that the electromagnetic coupling $e^2$ remains unchanged in cosmic time.

It is interesting to examine what the model predicts near the beginning of the universe, or $N_2 \approx 1$. Composite particles are described by $m_{n_A} \simeq \left(n_A + 2\right) m_P$ and $l_{n_A} \simeq \left(n_A + 2\right) l_P$. Thus, they have radii less than their Schwarzschild radii, i.e., they are black holes.

## X. VARIATION OF COUPLING CONSTANTS WITH COSMIC TIME

The dimensions of the Lagrangian (energy) and potential $(\mathbf{HQ})^2$ (length$^2$) imply that the constant multiplier of the Lagrangian has dimensions mass/time.We have expressed it as

$$m(N_2)\omega(N_2) \equiv N_2^{\,-1} m_P \omega_P \,. \tag{10.1}$$

The Lagrangian can be rewritten as

$$L = -\left[-m\omega^2\left(\mathbf{HQ})^2\right)\left(m\dot{\mathbf{Q}}^2\right) + m^2\omega^2\left(\dot{\mathbf{Q}}^{T}\cdot\mathbf{HQ}\right)^2\right], \tag{10.2}$$

allowing us to identify the familiar “harmonic-oscillator” coupling constant

$$k_{h.o.} \equiv m(N_2)\omega^2(N_2) = N_2^{\,-3/2} m_P \omega_P^{\,2} \,. \tag{10.3}$$

Thus, at the earliest cosmic epoch,

$$k_{h.o.} \simeq m_P \omega_P{}^2. \qquad (10.4)$$

The coupling is about $10^{-60}$ weaker today than it was in the earliest epoch. This factor is responsible, for example, for composite particle size and mass. As long as $\varepsilon$ is infinitesimal, the same forms of Veneziano-like amplitudes describe both nucleon and Planck-scale particle scattering.

The Lagrangian adopted in Article II, for virtual-particle exchange interactions, has the form

$$L = -\left\{-k\left[\mathbf{A}(\mathbf{Q})\right]^2 m\dot{\mathbf{Q}}^2 + mk\left[\dot{\mathbf{Q}}^T \cdot \mathbf{A}(\mathbf{Q})\right]^2\right\}^{1/2}. \qquad (10.5)$$

Here we are not concerned with the explicit form of the vector potential **A(Q),** but let us assume that it has dimension length$^{-1/2}$. Thus, k has dimensions mass-length$^3$-time$^{-2}$. Similarly to the procedure for the harmonic-oscillator potential, we assume that at the earliest epoch,

$$k(N_2 = 1) \simeq m_P l_P{}^3 \omega_P{}^2. \qquad (10.6)$$

If, at other epochs, the coupling is

$$k\left(N_2\right) \simeq m\left(N_2\right) l^3\left(N_2\right) \omega^2\left(N_2\right), \qquad (10.7)$$

then we can identify it as the electromagnetic coupling, or

$$k_{e.m.}(N_2) \simeq m_P l_P{}^3 \omega_P{}^2 = \hbar c \simeq e^2. \qquad (10.8)$$

Thus, in the model, the electromagnetic coupling remains independent of $N_2$, i.e., cosmic time.

Now, the coupling $k(N_2{=}1)$ can also be expressed as

$$k(N_2 = 1) \simeq m_P l_P{}^3 \omega_P{}^2 = m_P{}^2 G. \qquad (10.9)$$

By assuming that the gravitational constant G remains unchanged in cosmic time, let us again follow the same prescription for arbitrary $N_2$. The coupling becomes the gravitational coupling

$$k_{grav} = m^2(N_2)\, G. \qquad (10.10)$$

We thus have the ratio

$$\frac{k_{e.m.}}{k_{grav}} \approx N_2. \qquad (10.11)$$

This implies that in today's epoch, the gravitational coupling is about $10^{-40}$ weaker than the electromagnetic coupling, agreeing with experiment. At the earliest epoch, the couplings are equal.

## XI. EFFECT OF BACKGROUND INDEPENDENCE ON HADRON AND LEPTON SPINS

## A. Spin phase space

To introduce spin variables corresponding to the space-time variables, a second phase space is added to the phase space of the quark coordinates (see, e.g., Berezin and Marinov,[30] and Casalbuoni[31]). Let the Lagrangian L be a function of boson variables z and z* and fermion variables $\Theta$ and $\Theta^*$. The latter are Grassman variables which obey

$$\Theta^2 = \Theta^{*2} = \Theta^*\Theta + \Theta\Theta^* = 0. \tag{11.1}$$

Define the Hamiltonian function by the Legendre transformation

$$H\left(z, z^*; \Theta, \Theta^*\right) = i\frac{h}{2}\left(z^*\dot{z} - \dot{z}^* z\right) + i\frac{h}{2}\left(\Theta^*\dot{\Theta} - \dot{\Theta}^*\Theta\right) - L\left(z, \dot{z}, z^*, \dot{z}^*; \Theta, \dot{\Theta}, \Theta^*, \dot{\Theta}^*\right). \tag{11.2}$$

Introduce real variables

$$q = \sqrt{\frac{h}{2m\omega}}\left(z^* + z\right), \qquad p = \sqrt{\frac{hm\omega}{2}}\left(z^* - z\right),$$

$$q_F = \sqrt{\frac{h}{2m\omega}}\left(\Theta^* + \Theta\right), \qquad p_F = \sqrt{\frac{hm\omega}{2}}\left(\Theta^* - \Theta\right), \tag{11.3}$$

and re-express the Legendre transformation as

$$H\left(q, p; q_F, p_F\right) = \frac{1}{2}\left(p\dot{q} - q\dot{p}\right) + \frac{i}{2}\left(m\omega q_F \dot{q}_F + \frac{p_F \dot{p}_F}{m\omega}\right) - L\left(q, \dot{q}; q_F, \dot{q}_F, p_F, \dot{p}_F\right). \tag{11.4}$$

Finally, neglecting a total derivative and redefining fermi variables as

$$\xi_1 \equiv \sqrt{m\omega}\, q_F, \qquad \xi_2 = \frac{1}{\sqrt{m\omega}} p_F, \tag{11.5}$$

we obtain

$$H\left(q, p; \xi_1, \xi_2\right) = p\dot{q} + \frac{i}{2}\sum_{\alpha=1,2} \xi_\alpha \dot{\xi}_\alpha - L\left(q, \dot{q}; \xi_\alpha, \dot{\xi}_\alpha\right). \tag{11.6}$$

We shall use right-hand derivatives as defined by Casalbuoni.[31] The classical fermi variables obey

$$\left(\xi_{\alpha\mu}\right)^2 = \xi_{1\mu}\xi_{2\nu} + \xi_{2\upsilon}\xi_{1\mu} = 0. \tag{11.7}$$

The Lagrange equations are

$$\frac{d}{ds}\left(\frac{\partial L}{\partial \dot{\xi}_\alpha}\right) = \frac{\partial L}{\partial \xi_\alpha}, \; \alpha = 1, 2; \qquad \frac{d}{ds}\left(\frac{\partial L}{\partial \dot{q}}\right) = \frac{\partial L}{\partial q}. \tag{11.8}$$

Conjugate momenta are

$$\pi_\alpha = \frac{\partial L}{\partial \dot{\xi}_\alpha}; \qquad p = \frac{\partial L}{\partial \dot{q}}. \tag{11.9}$$

Thus, from the Lagrange equations,

$$\dot{\pi}_\alpha = \frac{\partial L}{\partial \xi_\alpha}; \quad \dot{p} = \frac{\partial L}{\partial q}. \tag{11.10}$$

Since H is a functions of $q, p; \xi_1, \xi_2$, we can write

$$dH = d\xi_1 \frac{\partial H}{\partial \xi_1} + d\xi_2 \frac{\partial H}{\partial \xi_2} + dq \frac{\partial H}{\partial q} + dp \frac{\partial H}{\partial p}. \tag{11.11}$$

On the other hand, we also can write

$$dH = \sum_{\alpha=1,2} d\xi_\alpha \left(\frac{i}{2}\dot{\xi}_\alpha - \dot{\pi}_\alpha\right) - \sum_{\alpha=1,2} d\dot{\xi}_\alpha \left(\frac{i}{2}\xi_\alpha - \pi_\alpha\right) + \dot{q}\cdot dp - \dot{p}\cdot dq. \tag{11.12}$$

Thus, we have the constraints

$$\pi_\alpha = -\frac{i}{2}\xi_\alpha, \tag{11.13}$$

so that

$$dH = i \sum_{\alpha=1,2} d\xi_\alpha \dot{\xi}_\alpha + \dot{q}dp - \dot{p}dq. \tag{11.14}$$

The Hamiltonian equations of motion are therefore

$$\dot{\xi}_\alpha = 2i \frac{\partial H}{\partial \xi_\alpha}; \qquad \dot{q} = \frac{\partial H}{\partial p}, \qquad \dot{p} = -\frac{\partial H}{\partial q}. \tag{11.15}$$

### B. Hadron-lepton model with spin

First, we introduce spin into the Lagrangian with a background frame. For simplicity, shorten the Lagrangian to $L = -\sqrt{-VT}$, which also yields the earlier Dirac Hamiltonians.

The system of 4N quarks are described by the space-time coordinates $Q_{IA}$ and spin coordinates $\Xi^\alpha_{IA}$, with I=1, 2, ..., N; A=1, 2, ,3 4; and $\alpha = 1, 2$. Consider the Lagrangian

$$L(s) = -m\omega\left[-V\left(\dot{\mathbf{Q}}^2(s) - \frac{i}{m\omega}\dot{\Xi}^1\cdot\dot{\Xi}^2\right)\right]^{1/2}, \tag{11.16}$$

where

$$V = \left(\mathbf{Q}(s)\cdot\mathbf{G}_4\mathbf{Q}(s) - \frac{i}{m\omega}\Xi^1\cdot\left(\mathbf{G}_4 - \mathbf{1}\right)\Xi^2\right). \tag{11.17}$$

and the $\Xi_\alpha$ obey the classical relations

$$\left(\Xi^{\alpha}_{IA\mu}\right)^2 = \Xi^{\alpha}_{IA\mu}\Xi^{\beta}_{JB\nu} + \Xi^{\beta}_{JB\nu}\Xi^{\alpha}_{IA\mu} = 0. \tag{11.18}$$

The conjugate momenta are

$$\mathbf{P} = \frac{\partial L}{\partial \dot{\mathbf{Q}}} = m\omega V\,\dot{\mathbf{Q}}\left[-V\left(\dot{\mathbf{Q}}^2 - \frac{i}{m\omega}\dot{\Xi}^1\cdot\dot{\Xi}^2\right)\right]^{-1/2}, \tag{11.19}$$

$$\Pi^1 = \frac{\partial L}{\partial \dot{\Xi}^1} = -\frac{i}{2}\dot{\Xi}^2\,V\left[-V\left(\dot{\mathbf{Q}}^2 - \frac{i}{m\omega}\dot{\Xi}^1\cdot\dot{\Xi}^2\right)\right]^{-1/2}, \tag{11.20}$$

$$\Pi^2 = \frac{\partial L}{\partial \dot{\Xi}^2} = \frac{i}{2}\dot{\Xi}^1\,V\left[-V\left(\dot{\mathbf{Q}}^2 - \frac{i}{m\omega}\dot{\Xi}^1\cdot\dot{\Xi}^2\right)\right]^{-1/2}. \tag{11.21}$$

Following the earlier discussion, we deduce that

$$\Pi^{\alpha} = -\frac{i}{2}\Xi^{\alpha}. \tag{11.22}$$

and arrive at the primary constraint

$$\Phi = \mathbf{P}^2 - im\omega\,\Xi^1\cdot\Xi^2 + m^2\omega^2 V = 0. \tag{11.23}$$

The Dirac Hamiltonian is taken to be

$$H = \frac{1}{2m}\left[\left(\mathbf{P}^2 + m^2\omega^2\mathbf{Q}\cdot\mathbf{G}_4\mathbf{Q}\right) - im\omega\left(\Xi^1\cdot\mathbf{G}_4\Xi^2\right)\right]. \tag{11.24}$$

In analogy to the transformation from $Q_{IA}$ to $y_{IA}$ and $W_A$, we define

$$\Xi^{\alpha}_{IA} \equiv \chi^{\alpha}_{IA} + N^{-1/2}\{\delta\,\vartheta^{\alpha}\}_A, \qquad \text{with } \sum_{I=1}^{N}\chi^{\alpha}_{IA} = 0. \tag{11.25}$$

The Hamiltonian can now be expressed as

$$H = \frac{1}{2m}\left[\sum_{I=1}^{N}\sum_{A=1}^{4} p_{IA}{}^2 + \sum_{A=1,3} p_A{}^2 + \sum_{A=2,4}\left(p_A{}^2 + m^2\omega^2 W_A{}^2\right) - im\omega\sum_{A=2,4}\vartheta^1_A\vartheta^2_A\right]. \tag{11.26}$$

The equations of motion for the spin variables are

$$\chi^{\alpha}_{IA} = \text{const.}, A = 1,2,3,4; \quad \vartheta^{\alpha}_A = \text{const.}, A = 1,3: \quad \ddot{\vartheta}^{\alpha}_A = -\omega^2\vartheta^{\alpha}_A, \quad A = 2,4. \tag{11.27}$$

Quantize the spin coordinates by applying the anticommutation relations

$$\{\chi^{\alpha}_{IA\mu}, \chi^{\beta}_{IA\nu}\} = -\hbar\, g_{\mu\nu}\delta_{\alpha\beta}; \qquad \{\vartheta^{\alpha}_{A\mu}, \vartheta^{\beta}_{A\nu}\} = -2N\hbar\, g_{\mu\nu}\delta_{\alpha\beta}. \tag{11.28}$$

For the constant operators, write the solutions as

$$\chi^{\alpha}_{IA\mu} = \sqrt{\hbar}\,\xi^{\alpha}_{IA\mu}, \quad A = 1,2,3,4; \quad \vartheta^{\alpha}_{A\mu} = \sqrt{2N\hbar}\,\zeta^{\alpha}_{A\mu}, \quad A = 1,3; \tag{11.29}$$

where, following Berezin and Marinov,[30] we introduce spin phase space operators

$\xi^{\alpha}_{IA\mu}, \xi^{\alpha}_{IA5}, \zeta^{\alpha}_{A\mu}, \zeta^{\alpha}_{A5}$ which obey anticommutation relations

$$\left\{\xi^{\alpha}_{IA\mu}, \xi^{\beta}_{IA\nu}\right\} = -g_{\mu\nu}\delta_{\alpha\beta}, \qquad \left\{\xi^{\alpha}_{IA5}, \xi^{\beta}_{IA5}\right\} = \delta_{\alpha\beta};$$

$$\left\{\zeta^{\alpha}_{A\mu}, \zeta^{\beta}_{A\nu}\right\} = -g_{\mu\nu}\delta_{\alpha\beta}, \qquad \left\{\zeta^{\alpha}_{A5}, \zeta^{\beta}_{A5}\right\} = \delta_{\alpha\beta}. \tag{11.30}$$

The two sets of generators each produce a Clifford algebra $C_5$, and are represented by Dirac gamma matrices as follows:

$$\xi^{\alpha}_{IA\mu} = \frac{1}{\sqrt{2}}\gamma^{\alpha}_{IA5}\gamma^{\alpha}_{IA\mu}, \quad \xi^{\alpha}_{IA5} = \frac{1}{\sqrt{2}}\gamma^{\alpha}_{IA5};$$
$$\zeta^{\alpha}_{A\mu} = \frac{1}{\sqrt{2}}\gamma^{\alpha}_{A5}\gamma^{\alpha}_{A\mu}, \quad \zeta^{\alpha}_{A5} = \frac{1}{\sqrt{2}}\gamma^{\alpha}_{A5}; \tag{11.31}$$

with

$$\left\{\gamma^{\alpha}_{IA\mu}, \gamma^{\beta}_{IA\nu}\right\} = 2g_{\mu\nu}\delta_{\alpha\beta}, \qquad \gamma^{\alpha}_{IA5} = i\gamma^{\alpha}_{IA0}\gamma^{\alpha}_{IA1}\gamma^{\alpha}_{IA2}\gamma^{\alpha}_{IA3};$$
$$\left\{\gamma^{\alpha}_{A\mu}, \gamma^{\beta}_{A\nu}\right\} = 2g_{\mu\nu}\delta_{\alpha\beta}, \qquad \gamma^{\alpha}_{A5} = i\gamma^{\alpha}_{A0}\gamma^{\alpha}_{A1}\gamma^{\alpha}_{A2}\gamma^{\alpha}_{A3}. \tag{11.32}$$

Write the remaining spin coordinate solutions as

$$\zeta^{2}_{A} = -\omega\zeta^{1}_{A}, \quad A = 2,4; \quad \zeta^{1}_{A} = (N\hbar)^{1/2}\left[b_{A}{}^{\dagger}\exp(i\omega s) + b_{A}\exp(-i\omega s)\right]; \tag{11.33}$$

where

$$\{b_{A\mu}, b_{A\nu}\} = \{b_{A\mu}{}^{\dagger}, b_{A\nu}{}^{\dagger}\} = 0; \;\; \{b_{A\mu}, b_{A\nu}{}^{\dagger}\} = -g_{\mu\nu}; A = 2,4. \tag{11.34}$$

The operator $\zeta^{1}_{A}$ yields integral spin while the $\xi^{\alpha}_{IA}$ and $\xi^{\alpha}_{A}$ yield half integral spin.

The Hamiltonian can now be expressed as

$$H = \frac{1}{2m}\left[\sum_{I=1}^{N}\sum_{A=1}^{4} p_{IA}{}^{2} + \sum_{A=1,3} p_{A}{}^{2} + 4N\hbar m\omega \sum_{A=2,4}\left(a_{A}{}^{\dagger}\cdot a_{A} + b_{A}{}^{\dagger}\cdot b_{A}\right)\right]. \tag{11.35}$$

The eigenstates of the Hamiltonian can be taken to be simultaneous eigenstates of the operators $a_{A}{}^{\dagger}a_{A}$ and $b_{A}{}^{\dagger}b_{A}$ , or of the operators $N_A$ and $\overline{N}_A$ defined below:

$$N_{A} \equiv -\left(a^{\dagger}_{A}a_{A} + b_{A}{}^{\dagger}b_{A}\right), \quad \overline{N}_{A} \equiv a^{\dagger}_{A}b_{A} + b_{A}{}^{\dagger}a_{A}, \quad A = 2,4. \tag{11.36}$$

Note that $\overline{N}_{A}{}^{2} = -N_{A}$.

The n.b.c. yield the following conditions on the classical variables

$$\left(P_{IA}\right)^{2} = 0, \; s \sim \pm\infty; \qquad \left(\Pi^{\alpha}_{IA}\right)^{2} \propto \left(\Xi^{\alpha}_{IA}\right)^{2} = 0, \; s \to \pm\infty. \tag{11.37}$$

The second condition is automatically satisfied by the properties of the Grassman variables.

One finds

$$\left(P_{IA}\right)^2 = \left(\Xi^1_{IA} \cdot P_{IA}\right)^2. \tag{11.38}$$

Since this relation is the square of a scalar, the n.b.c. imply that the following condition must also be satisfied:

$$\Xi^1_{IA} \cdot P_{IA} = 0, \quad s \to \pm\infty. \tag{11.39}$$

This equation yields a set of constraints. Using a spinor representation, we can express these, for the lepton quarks, as

$$\left(\gamma_{IA} \cdot p_{IA} + \gamma_3 \cdot p_3 - \overline{N}_4 mc\right) u_{IA}(p_{IA})\, u_3\left(p_3\right) \left|p_{IA}, p_3, N_4, \overline{N}_4\right\rangle = 0,$$

$$b_4 \cdot a_4 \left|a_{IA}, N_4, \overline{N}_4\right\rangle = 0,$$

$$\left\{\left[mc\ \left(\gamma_{IA} - \gamma_3\right) - p_{IA}\right] \cdot a_4\right\} u_{IA}(p_{IA})\, u_3\left(p_3\right) \left|p_{IA}, p_3, N_4, \overline{N}_4\right\rangle = 0, \ A = 3{,}4, \tag{11.40}$$

where the $u_{IA}$ and $u_3$ are four-component spinors. An analogous set of equations hold for the hadrons. Note that both sets of equations also describe a two-quark composite.

The first of the constraint equations above represents a generalized Dirac equation which is similar to the generalized Dirac equation constructed by Ramond in Ref. 32. The presence of two gamma functions (and therefore two spinors) implies that the corresponding composite has integer spin. The same equation yields the composite mass spectrum

$$\left(p_{IA} + p_3\right)^2 = N_4 m^2 c^2. \tag{11.41}$$

Similar remarks hold for the hadron composites.

To remove the background frame, the potential is modified to read

$$V = \left(\mathbf{Q}(s) \cdot \mathbf{HQ}(s) - \frac{i}{m\omega} \Xi^1 \cdot (\mathbf{H} - \mathbf{1})\, \Xi^2\right\}, \tag{11.42}$$

where the matrix **H** was defined in Sec. VI. The Dirac Hamiltonian is now

$$H = (1/2m)\left[\mathbf{P}^2 + m^2\omega^2 (\mathbf{HQ})^2 - im\omega\, \Xi^1 \cdot \mathbf{H}\Xi^2\right]. \tag{11.43}$$

Using the transformation (11.33), we obtain the Hamiltonian

$$H = \frac{1}{2m}\left\{\sum_{I=1}^{N} \sum_{A=1}^{4} \left[p_{IA}{}^2 - (\varepsilon m\omega)^2 y_{IA}{}^2\right] + p_3{}^2 + \left[p_1{}^2 - (\varepsilon m\omega)^2 W_1{}^2\right] + \sum_{A=2,4} \left[p_A{}^2 + (1 + i\varepsilon)^2 (m\omega)^2 W_A{}^2\right] + \varepsilon\omega \sum_I \sum_A \chi^1_{IA} \cdot \chi^2_{IA} + \varepsilon\omega\, \vartheta^1_3 \cdot \vartheta^2_3 + i\,(1 + i\varepsilon)^2 m\omega \sum_{A=2,4} \vartheta^1_A \vartheta^2_A\right\}. \tag{11.44}$$

The equations of motion for the fermion variables are

$$\ddot{\chi}^{\alpha}_{IA} = (\varepsilon\omega)^2 \chi^{\alpha}_{IA}; \quad \vartheta^{\alpha}_{1} = (\varepsilon\omega)^2 \vartheta^{\alpha}_{1}; \quad \vartheta^{\alpha}_{3} = \text{const.};$$

$$\ddot{\vartheta}^{\alpha}_{A} = -(1 + i\varepsilon)^2 \omega^2 \vartheta^{\alpha}_{A}, \quad A = 2, 4. \tag{11.45}$$

Thus, we see the spin variables $\gamma^{\alpha}_{3}$ as well as the variables $W_3$ and $p_3$ do not contribute to asymptotic solutions and therefore not to the lepton composites. As a result, the lepton composites have half-integer spin, while the two-quark hadron composites have integer spin. In the next section, we show that when a SU(3) symmetry is added, the hadrons split into baryons with half integer spin and meson with integer spin. The leptons retain their half-integer spin.

## XII. BARYONS, MESONS AND LEPTONS

We modify the hadron Dirac Hamiltonian of Sec. IV by introducing a new symmetry. For quark coordinates $Q^{C}_{IA}$, I=1, 2, . . . N, C=1, 2, 3, and A=1, 2, write

$$H = \frac{1}{2m}\left[\mathbf{P}^{T}\cdot\mathbf{P} + (m\omega)^2 \mathbf{Q}^{T}\cdot\mathbf{G}^2\mathbf{Q}\right] \approx 0, \tag{12.1}$$

where

$$\mathbf{G}_{3}{}^{2} \equiv \mathbf{g}_{2}{}^{2} \otimes \mathbf{g}_{SU3}{}^{2} \otimes \mathbf{N}^2, \tag{12.2}$$

and

$$\mathbf{g}_{SU3} \equiv \begin{pmatrix} g^{3}_{11} & g^{3}_{12} & g^{3}_{12} \\ g^{3}_{12} & g^{3}_{11} & g^{3}_{12} \\ g^{3}_{12} & g^{3}_{12} & g^{3}_{11} \end{pmatrix}. \tag{12.3}$$

We shall restrict consideration to cases $\mathbf{g}_{SU3}=\mathbf{g}_{SU3}{}^2$. Label the solutions as

$$\mathbf{g}_{3} \equiv \frac{1}{3}\begin{pmatrix} 2 & -1 & -1 \\ -1 & 2 & -1 \\ -1 & -1 & 2 \end{pmatrix}, \qquad \overline{\mathbf{g}}_{3} = \frac{1}{3}\begin{pmatrix} 1 & 1 & 1 \\ 1 & 1 & 1 \\ 1 & 1 & 1 \end{pmatrix}, \tag{12.4}$$

and the unit matrix $\mathbf{1}$. Note that $\mathbf{g}_3$ and $\overline{\mathbf{g}}_3$ are orthogonal, and $\mathbf{g}_3 + \overline{\mathbf{g}}_3 = \mathbf{1}$.

Make a transformation of coordinates:

$$\mathbf{Q}_{I} = \mathbf{y}_{I} + N^{-1/2}\left(\delta_2 \otimes \lambda\right)\mathbf{W}, \tag{12.5}$$

with

$$\lambda \equiv \begin{pmatrix} 1/3 & 1/3 & 1/2 \\ 1/3 & 1/3 & -1/2 \\ 1/3 & -2/3 & 0 \end{pmatrix}. \tag{12.6}$$

The matrix $\lambda$ diagonalizes both coupling matrices $\mathbf{g}_3$ and $\overline{\mathbf{g}}_3$ .

The momenta conjugate to W are

$$\mathbf{P_I} = \mathbf{p_I} + N^{-1/2}\left(\delta_2^{\ -1}\right)^T \otimes \left(\lambda^{-1}\right)^T \mathbf{p} = \mathbf{p}_I + N^{-1/2}\delta_2 \otimes \left(\lambda^{-1}\right)^T \mathbf{p}, \tag{12.7}$$

where

$$\left(\lambda^{-1}\right)^T = \begin{pmatrix} 1 & 1/2 & 1 \\ 1 & 1/2 & -1 \\ 1 & -1 & 0 \end{pmatrix}. \tag{12.8}$$

The hadron Hamiltonian becomes

$$H = \frac{1}{2m}\left\{\sum_{I=1}^{N}\sum_{A} p_{IA}{}^2 + 3\left(p_1^1\right)^2 + \frac{3}{2}\left(p_1^2\right)^2 + 2\left(p_1^3\right)^2 + \left[3\left(p_2^1\right)^2 + \frac{1}{3}(m\omega)^2\left(g_{11}^3 + 2g_{12}^3\right)\left(W_2^1\right)^2\right]\right.$$

$$\left. + \left[\frac{3}{2}\left(p_2^2\right)^2 + \frac{2}{3}(m\omega)^2\left(g_{11}^3 - g_{12}^3\right)\left(W_2^2\right)^2\right] + \left[2\left(p_2^2\right)^2 + \frac{1}{2}(m\omega)^2\left(g_{11}^3 - g_{12}^3\right)\left(W_2^3\right)^2\right]\right\}. \tag{12.9}$$

The equations of motion are

$$\dot{y}_{IA}^C = 0; \ \ddot{W}_1^C = 0,$$

$$\ddot{W}_2^1 = -\left(g_{11}^3 + 2g_{12}^3\right)\omega^2 W_2^1, \qquad \ddot{W}_2^C = -\left(g_{11}^3 - g_{12}^3\right)\omega^2 W_2^C, \ C = 2,3. \tag{12.10}$$

Writing out the quarks coordinates, we have,

$$Q_{IA}^1 = y_{IA}^1 + \frac{1}{\sqrt{2N}}\left\{\lambda W_1\right\}^1 \pm \frac{1}{\sqrt{2N}}\left[\frac{1}{3}W_2^1 + \frac{1}{3}W_2^2 + \frac{1}{2}W_2^3\right],$$

$$Q_{IA}^2 = y_{IA}^2 + \frac{1}{\sqrt{2N}}\left\{\lambda W_1\right\}^2 \pm \frac{1}{\sqrt{2N}}\left[\frac{1}{3}W_2^1 + \frac{1}{3}W_2^2 - \frac{1}{2}W_2^3\right],$$

$$Q_{IA}^3 = y_{IA}^3 + \frac{1}{\sqrt{2N}}\left\{\lambda W_1\right\}^3 \pm \frac{1}{\sqrt{2N}}\left[\frac{1}{3}W_2^1 - \frac{2}{3}W_2^2\right]. \tag{12.11}$$

For the hadron quarks, A=1, 2, we shall take $\mathbf{g}_{SU3}$=$\mathbf{g}_3$, or $g_{11}^3 = 2/3$, and $g_{12}^3 = 1/3 g_{12}$. The only oscillatory terms are solutions of

$$\ddot{W}_2^C = -\omega^2 W_2^C, \quad C = 2,3. \tag{12.12}$$

It is thus possible, with constraints on allowed values of $n_2^C$, to form three-quark composites and anticomposites which we designate as baryons and antibaryons, respectively:

$$\frac{1}{3}\sum_{C=1}^{3} Q_{I1}^C, \quad \text{and} \quad \frac{1}{3}\sum_{C=1}^{3} Q_{I2}^C. \tag{12.13}$$

Two-quark composites, designated as mesons, can be constructed from quark-antiquark pairs

$$\frac{1}{2}\left(Q_{I1}^C + Q_{I2}^C\right). \tag{12 .14}$$

The mass spectrums are different for the baryons and mesons.

For the lepton quarks, we carry out the similar formulation with $\mathbf{g}_{SU3} = \mathbf{1}$, which introduces no extra symmetry. We omit the symmetry label C. The oscillatory term is the solution to

$$\ddot{W}_4 = -\omega^2 W_4, \tag{12.15}$$

and only quark-antiquark leptons can be formed.

Thus the hadrons break up into two categories, baryons and mesons, while the leptons remain a single category. The total Hamiltonian for all the quark coordinates is

$$H = \frac{1}{2m}\left[\mathbf{P}^T \cdot \mathbf{P} + (m\omega)^2 \mathbf{Q}^T \cdot \mathbf{G}_{43}{}^2 \mathbf{Q}\right] \approx 0, \tag{12.16}$$

with

$$\mathbf{G}_{43} \equiv \mathbf{g}_{43} \otimes \mathbf{N}, \text{ with } \mathbf{g}_{43} \equiv \begin{pmatrix} \mathbf{g}_3 \otimes \mathbf{g}_2 & 0 \\ 0 & \mathbf{1} \otimes \mathbf{g}_2 \end{pmatrix}. \tag{12.17}$$

To eliminate the background reference frame, redefine the harmonic-oscillator potential as

$$-\frac{1}{2N}\sum_{I,J}^{N}\sum_{A,B}^{4}\sum_{C,D}^{3}\left(\mathbf{h}_{43}{}^2\right)_{AB}^{CD}\left[x_{IA}^C(s) - x_{JB}^D(s)\right]^2, \tag{12.18}$$

where $\mathbf{h}_{43}$ is

$$\mathbf{h}_{43} \equiv \sqrt{1+2i\varepsilon}\begin{pmatrix} \mathbf{g}_3 \otimes \mathbf{g}_2 + \frac{\varepsilon}{2}\overline{\mathbf{g}}_3 \otimes \overline{\mathbf{g}_2} & \frac{\varepsilon}{2}\overline{\mathbf{g}}_3 \otimes \overline{\mathbf{g}_2} \\ \frac{\varepsilon}{2}\overline{\mathbf{g}}_3 \otimes \overline{\mathbf{g}_2} & \mathbf{1} \otimes \mathbf{g}_2 + \frac{\varepsilon}{2}\overline{\mathbf{g}}_3 \otimes \overline{\mathbf{g}_2} \end{pmatrix}. \tag{12.19}$$

Transforming back to the physical coordinates $\mathbf{Q}$, we find

$$H = \frac{1}{2m}\left[\mathbf{P}^T \cdot \mathbf{P} + (m\omega)^2 \mathbf{Q}^T \cdot \mathbf{H}_{34}{}^2 \mathbf{Q}\right] \approx 0, \tag{12.20}$$

where

$$\mathbf{H}_{43} \equiv \mathbf{g}_{43} \otimes \left(\mathbf{g}_4 - i\varepsilon \mathbf{f}_4\right) \otimes \mathbf{N} + i\varepsilon\, \mathbf{1} \otimes \mathbf{1}. \tag{12.21}$$

The center-of-mass coordinate $W_3$ (for the generalized coordinates $x_{IA}$) is again the only variable linear in s, and thus does not contribute to composite formulations in the asymptotic regions. With the spin calculations in Sec. II, it is straight forward to show that the baryons and leptons have half-integer spin, while the mesons have integer spin, in agreement with experiment.

## XIII. A DETAILED SUMMARY

To obtain a relativistic quantum model independent of background reference frame, we begin with a parametrically invariant action with a single evolution parameter s independent of space-time dynamics. This yields a nonstandard Lagrangian. The variational principle leads to natural boundary conditions (n.b.c.) that are invariant under Lorentz transformations but not under contact transformations. Thus, a distinction is made between physical coordinates and generalized coordinates.

First, we look at a simple relativistic harmonic-oscillator Lagrangian describing a system of quarks *with* a background frame. Spin and internal symmetries are omitted. Input parameters are the Planck units and the cosmic number $N_2$. All quarks oscillate with the same frequency and amplitude about straight-line trajectories in space-time. Pairs of quarks form composite particles by physical initial conditions. The n.b.c. yield mass-shell relations and suppress unwanted time-oscillations. However, the n.b.c. together with the Dirac constraints imply the composites have zero mass.

The coupling matrix is then modified by including an infinitesimal positive parameter $\varepsilon$. Although this simplest version of the model doesn't describe massive composite particles, it yields quark solutions representing harmonic oscillations about trajectories determined by gravity in the flat Rindler space-time near the horizon of a spherically symmetric static black hole.

The zero-mass difficulty is removed by considering two identical systems of quarks (i..e., introducing a "supersymmetry"), yielding the 4N quarks $Q_{IA}$, I=1, 2,..., N, and A=1, 2, 3, 4. Before modifying the harmonic-oscillator coupling matrix to include $\varepsilon$, we look at two 4x4 diagonalizing matrices, $\delta$ and $\rho$. The matrix $\delta$ leads to solutions symmetric under the interchange of lepton and hadron quark coordinates. The center-of-mass vector involves both hadron and lepton quarks. Solutions corresponding to imaginary-mass leptons allow real-mass hadrons, and vice-versa. Free leptons and hadrons result from initial conditions on pairs of quarks.

A Lagrangian of the same form is expressed in terms of generalized coordinates $x_{IA}$. The diagonalizing matrix $\rho$ is chosen, yielding solutions asymmetric under the interchange of the two systems. A suitable definition of the lepton and hadron *physical* coordinates in terms of these generalized coordinates reproduces the earlier quark solutions. The c.m. vector for the generalized coordinates involves only a *lepton* variable. This asymmetry takes on importance in the final model, where the c.m. coordinate and its conjugate momentum play no role.

To eliminate the background frame, the positive infinitesimal parameter $\varepsilon$ is included such that the Lagrangian in generalized coordinates retains the lepton-hadron symmetry. The matrix $\delta$ no longer diagonalizes the coupling matrix, and we must use $\rho$. Physical quark coordinates are defined in terms of the generalized coordinates, and the lepton-hadron symmetry is spontaneously broken.

In order to solve the equations of motion, a transformation is made to a set of 4N+4 coordinates which are reduced to 4N coordinates by constraints. Under certain conditions cluster decomposition is possible. The clusters considered in this paper are limited to four quarks. Quark confinement and elastic scattering of composite particles result from a combination of initial conditions on two of the quarks and cluster decomposition. Real mass composites in the initial state imply real mass composites in the final state. There is no coupling between real mass particles and imaginary mass particle, or between particles and separate quarks. The cluster space-time is not observable, acting like a black hole absorbing and emitting particles.

The vanishing Hamiltonian implies that the state vector $|\Psi\rangle$, which contains information for both $t = \pm\infty$, is independent of s. Since $|\Psi\rangle$ cannot be completely specified by physical initial conditions at $t = -\infty$, the initial state is a sum over all possible states satisfying these conditions. The final state is expressed as a sum over the complete set of eigenstates of the commuting operators including H. There is no “S-Matrix.” The scattering amplitude is the overlap of the two states, implying the final state is a physical state describing real-mass particles. There are no unitarity or causality problems.

Lepton composites scatter only in the forward or backward directions, which, for these identical particles, is tantamount to no scattering at all. The hadron-hadron scattering amplitude takes a form characteristic of the Veneziano amplitude and other dual models. Poles in the amplitude can be interpreted as resonances.

The internal angular momentum was not specified in the scattering amplitude calculations. However, since it is involved in resonance production, the complete specification of states is carried out. Contrary to the nonrelativistic harmonic oscillator, there is no degeneracy in mass for a given angular momentum, and $l_A = n_A$. For this spinless model, the excitations of each composite lies on a single Regge trajectory.

The input parameter $N_2=(L_H/l_P)^{2/3}$ leaves the cosmic-number relation $N_1=N_2$ permanent as $N_2$ changes. The expansion of the observable universe implies not only an increasing number of quarks, but particles decreasing in mass and increasing in size. However, in a model of similar form, the electromagnetic coupling remains unchanged, so that, for example, we would expect no change in the crystalline structure of a rock.

At this moment in cosmic time, cosmologists estimate that $N_2 \simeq 10^{40}$. In this simple model, this yields composite particles of nucleon mass and size. An observer on earth finds that particle masses on Star X are the same as on earth. At epochs when $N_2$ is near unity, the composite radius is less than its Schwarzschild radius. The electromagnetic and gravitational couplings are equal at the epoch $N_2=1$. Hadron composites scatter via Veneziano-type amplitudes at all epochs of an observable universe as long as $\varepsilon$ is infinitesimal.

A brief consideration of the addition of spin and SU(3) internal symmetry to the model shows that the elimination of the background frame can explain a further experimental result. Hadrons come as three-quark baryons of half-integer spin or quark-antiquark mesons with integer spin, while the leptons remain quark-antiquark composites of half integer spin.

## XIV. DISCUSSION

The prevailing theory in particle physics is the very successful Standard Model which describes all known interactions except gravity. However, it is an effective theory depending on a large number of parameters. Further, to be complete, it ought to include quantum gravity to describe elementary particle scattering at very short distances. Thus it has been generally believed there must exist a more fundamental theory from which these parameters, many of which are masses, can be deduced. Two other important approaches in particle physics are Superstrings and Supersymmetry. Superstrings describe elementary particles as strings with sizes on the order of the Planck length, that is, the physics takes place in regions of space-time not testable by present experimental technology. Supersymmetry predicts the existence of particles not yet observed. Nevertheless, the two theories have provided clues towards a cosmological theory which includes quantum gravity.

Quantum gravity is also the goal of researchers who have believed that the elimination of a background reference frame in quantum theory is a necessary ingredient.[14,15,16] This elimination has not been accomplished for any of the three theories above. The main purpose of this article has been to explicitly construct a quantum model that does. The results have gone far beyond the original goal, providing a quantum unification of gravitation and the harmonic-oscillator force near the horizons of microscopic black holes.

Although the model is based on a particle ontology, we can identify some common ground with effective field theories. The Lagrangian is characterized by a gauge invariance. A “Higgs” mechanism is necessary to give the particles mass. The Lagrangian exhibits a “spontaneously broken” symmetry, such that the solutions, as a class, do not exhibit the symmetry of the Lagrangian. Under a transformation of coordinates, the symmetry becomes “hidden” while composite-particle mass shell constraints $p^2=m^2c^2$ are “revealed.”

Superstring theory has provided insights that suggest the unification of forces and the possible inclusion of quantum gravity. Yet, with Planck-length strings, it is not testable experimentally. More disturbing is that extra dimensions are introduced not because of physical or philosophical motivation, but to have a mathematically consistent theory, namely one where the constraint algebra can be closed.[33,34] In the model of this article, the constraint algebra is closed, and the formulation is in four dimensions. It is interesting that we can relate complications of Superstring theory to similar ones in field theory. Both strings and fields are physically extended objects. Both can be thought of as an infinite number of beads on an imaginary string. This produces difficulties because of separate modes of oscillation for each “bead.”

Other new results of the model here include the cosmological determination of realistic masses and the prediction of a lepton substructure that does not contradict present experimental evidence. This does not negate the possibility of experiments which directly or indirectly confirm the substructure. For example, in Article II, quark substructure allows a straight forward calculation of Feynman diagrams for the exchange of virtual particles in composite elastic and inelastic scattering. States of the present model are used as the initial states in the perturbation theory.

Also new is the prediction that quarks are confined by decoupling constraints rather than a binding force. Interestingly, as shown in Article II, particle-exchange forces also do not result in forces between confined quarks of the composite particles. The quarks thus remain confined by decoupling conditions alone.

Finally, it is significant that this particle model avoids the difficulties usually associated with action-at-a-distance theories and particle ontologies. It is generally thought that local quantum field theory is the only way to combine a quantum theory of particles with special relativity and maintain causality.[35,36] This is not true, as further demonstrated by the effective model for particle-exchange interactions in Article II, where Feynman propagators are derived.

## XV. DISCUSSION OF THE PARTICLE ONTOLOGY

The relativistic quantized model is based on a particle ontology leading to a simple generalization of nonrelativistic quantum mechanics. In the past, attempts to formulate relativistic quantum particle theories were abandoned because of unresolved difficulties and concomitant successes in field theory. Although a good deal of attention has been given to the conceptual foundations of field theory,[37,38] very little consideration attends to relativistic particle ontologies.[39] It has been argued by S. Weinberg[36] that, with some caveats, "quantum mechanics plus Lorentz invariance plus cluster decomposition implies quantum field theory." One of those caveats is that the Hamiltonian has a decomposition $H=H_0+H_I$ such that $H_0$ generates the unperturbed equations of motion.. The models of this paper and the next have no such decomposition.

Although the use of quantum fields has become a dominant approach, field theory did not immediately permeate relativistic particle physics. Difficulties, such as infinite self-energies, kept alive attempts to formulate particle ontologies throughout the middle decades of the 20th century.[39] One of the best known is the classical action-at-a-distance model of electrodynamics formulated by J. A. Wheeler and R. P. Feynman.[40,41] However, until now, all efforts fell short, resulting, for example, in particles exceeding the velocity of light, lack of quantization, lack of cluster decomposition, and no particle creation and annihilation. Meanwhile, the difficulties in field theory were overcome, and the approach culminated in the formulation of the Standard Model.